\documentclass[%
 reprint,
superscriptaddress,
 amsmath,amssymb,
 aps,
prstab,
]{revtex4-1}

\usepackage{graphicx}
\usepackage{dcolumn}
\usepackage{bm}
\usepackage{amsmath}

\begin{document}


\title{High-brilliance, high-flux compact inverse Compton light source}

\author{K. E. Deitrick}
\email{kdeit001@odu.edu}
\altaffiliation{now at Jefferson Lab}
\email{deitrick@jlab.org}
\affiliation{Department of Physics, Center for Accelerator Science, Old Dominion University, Norfolk, Virginia 23529, USA}
\author{G. A. Krafft}
\affiliation{Department of Physics, Center for Accelerator Science, Old Dominion University, Norfolk, Virginia 23529, USA}
\affiliation{Thomas Jefferson National Accelerator Facility, Newport News, Virginia 23606, USA}
\author{B. Terzi\'{c}}
\affiliation{Department of Physics, Center for Accelerator Science, Old Dominion University, Norfolk, Virginia 23529, USA}
\author{J. R. Delayen}
\affiliation{Department of Physics, Center for Accelerator Science, Old Dominion University, Norfolk, Virginia 23529, USA}
\affiliation{Thomas Jefferson National Accelerator Facility, Newport News, Virginia 23606, USA}

\date{\today}

\begin{abstract}

A compact Inverse Compton Light Source (ICLS) design is presented, with flux and brilliance orders of magnitude beyond conventional laboratory-scale sources and other compact ICLS designs. This design utilizes the physics of inverse Compton scattering of an extremely low emittance electron beam by a laser pulse of \textit{rms} length of approximately two-thirds of a picosecond (2/3~ps). The accelerator is composed of a superconducting radiofrequency (SRF) reentrant gun followed by four double-spoke SRF cavities. After the linac are three quadrupole magnets to focus the electron beam to the interaction point (IP). The distance from cathode surface to IP is less than 6 meters, with the cathode producing electron bunches with a bunch charge of 10~pC and a few picoseconds in length. The incident laser has 1~MW circulating power, a 1~micron wavelength, and a spot size of 3.2~microns at the IP. The repetition rate of this source is 100~MHz, in order to achieve a high flux despite the low bunch charge. The anticipated X-ray source parameters include an energy of 12~keV, with a total flux of $1.4\times10^{14}$~ph/s, the flux into a 0.1\% bandwidth of $2.1\times10^{11}$~ph/(s-0.1\%BW), and the average brilliance of $2.2\times10^{15}$~ph/(s-mm$^2$-mrad$^2$-0.1\%BW).

\end{abstract}

\maketitle


\section{\label{SEC:INTRO}Introduction}

Since their discovery in 1895, X-rays have been a powerful technique for determining the structure of condensed matter. For the first 70 years of using X-rays, sources barely changed from the original bremsstrahlung tubes used in their discovery~\cite{XrayTube}. Until recently, large accelerator-based synchrotron facilities set the standard for the highest quality X-ray beams. At present, this standard has been largely surpassed in free electron lasers (FELs).

Most high-brilliance sources exist at large facilities, especially third-generation synchrotrons~\cite{XRay_Couprie}. However, due to various concerns, among them cost, risk of transporting valuable items, and limited available runtime at large facilities, there has been an increasing demand for laboratory-scale sources. Sometimes referred to as ``compact'', one description is any machine that fits in a 100 m$^2$ area. Additional desirable constraints are that the purchase and operating cost are not prohibitive for the smaller facilities and that the operation of a such a machine is possible by non-experts.

There are many X-ray experimental techniques that exist today; any given technique may be utilized in a wide range of fields. Some of the more prominent techniques currently in use include phase contrast imaging (PCI), absorption radiography, K-edge subtraction imaging, radiotherapy (treatment of tumors with X-rays), and computed tomography (CT). Some of the fields in which these techniques are used include medicine, cultural heritage, material science development, and industry~\cite{NIMB_Jacquet,RAST_Krafft}. Given the wide range of applications, the increasing demand for higher quality X-ray sources is understandable. In this paper, we present a design of a compact Inverse Compton Source based on SRF beam acceleration which was outlined in \cite{IPAC2017}. Because the SRF is run continuous wave (CW), high average flux and brilliance are possible.

\begin{figure}
\includegraphics[width=\columnwidth]{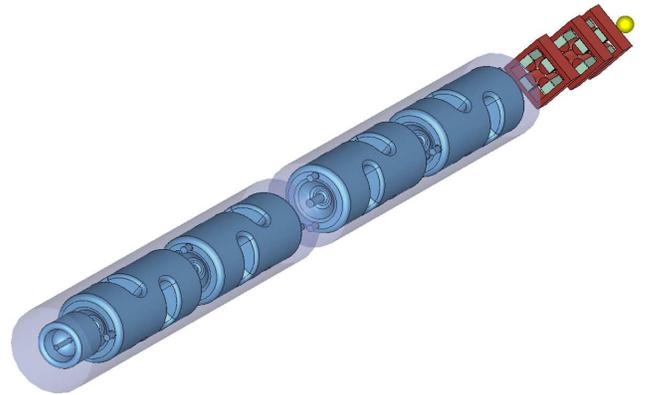}
\caption{A schematic of the entire design. The first cryomodule contains the gun and two double-spoke cavities, the second contains the last two double-spoke cavities. Three quadrupole magnets (red) follow the linac, before the interaction point (yellow). }
\label{FIG:ENTIRE}
\end{figure}

This paper is organized as follows. A brief overview of relevant SRF, electron beam, and X-ray beam parameters is presented in Sec.~\ref{SEC:PARAS}. Sec.~\ref{SEC:ICLS} goes into detail about Inverse Compton Light Sources, inverse Compton scattering, and compact ICLS designs. Our design consists of three separate regions: the SRF reentrant gun (Sec.~\ref{SEC:GUN}), the SRF linear accelerator (Sec.~\ref{SEC:LINAC}), and the final focusing (Sec.~\ref{SEC:FOCUSING}). A complete layout of these design components is shown in Fig.~\ref{FIG:ENTIRE}. In Sec.~\ref{SEC:JITTER} the sensitivity study results are presented, while the incident laser is addressed in Sec.~\ref{SEC:LASER}. The complete X-ray beam parameters are presented in Sec.~\ref{SEC:XRAY}, plans for future work are given in Sec.~\ref{SEC:FURTHER}, before the final summary presented in Sec.~\ref{SEC:CONCLUSION}. The design presented in this paper potentially outperforms all other compact ICLS designs.

\section{\label{SEC:PARAS}Beam Parameters}

\subsection{Electron Beam Parameters}

In our simulations, each particle in the beam is described by a set of six coordinates: ($x$, $p_x$, $y$, $p_y$, $z$, $p_z$) where $x$ and $y$ are the transverse positions of the particle, $p_x$ and $p_y$ are the transverse momenta, $z$ is the longitudinal position relative to a reference particle, and $p_z$ is the momentum along the beam trajectory. For a free particle, the energy $E$ of any particle within the bunch is related to its total momentum $p$ by $\beta E = cp = c\sqrt{p_x^2 + p_y^2 + p_z^2}$.

Following standard practices, it is often more convenient to use an alternate set of coordinates: ($x$, $x'$, $y$, $y'$, $z$, $\delta$) where $x' \equiv p_x/p_z$, $y' \equiv p_y/p_z$, and $\delta \equiv \Delta p/p_0$ such that $\Delta p \equiv p - p_0$, with $p_0$ representing the momentum of a particle with the average energy of the bunch. When the relative momentum error $\delta$ is small, $x' \approx p_x/\langle p_z\rangle$.

In this paper, beam sizes are quoted using the root mean square (\textit{rms}) of the particle distribution. The unnormalized Sacherer \textit{rms} emittance is defined by
\begin{equation}
\label{EQ:SACHERER}
\epsilon_{x,\mathit{rms}} = \sqrt{\sigma_x^2\sigma_{x'}^2 - \sigma_{xx'}^2}
\end{equation}
with $\sigma_x \equiv \sqrt{\langle x^2\rangle - \langle x\rangle^2}$, $\sigma_{x'} \equiv \sqrt{\langle x'^2\rangle - \langle x'\rangle^2}$, and $\sigma_{xx'} \equiv  \langle xx'\rangle - \langle x\rangle\langle x'\rangle$. We also use the normalized emittances given by
\begin{equation}
\label{EQ:EMIT_RELATION}
\epsilon^N_{x,\mathit{rms}} = \beta\gamma\epsilon_{x,\mathit{rms}}.
\end{equation}

\subsection{\label{SEC:XRAY_PARAS}X-ray Beam Parameters}

The total flux of a photon beam, $\mathcal{F}$, is the rate at which the photons pass a given location with units of photons/sec. The formula specific to a photon beam produced by inverse Compton scattering will be given in Sec.~\ref{SEC:COMPTON_SCATTERING}. The parameter $\mathcal{F}_{0.1\%}$ represents the flux in a 0.1\% bandwidth.

The spectral brightness or brilliance of a photon beam is the density of photons in the six-dimensional space containing the beam. The general formula for the brilliance of a photon beam into a 0.1\% bandwidth is
\begin{equation}
\label{EQ:GEN_BRILLIANCE}
\mathcal{B} = \frac{\mathcal{F}_{0.1\%}}{4\pi^2\sigma_{\gamma,x}\sigma_{\gamma,x'}\sigma_{\gamma,y}\sigma_{\gamma,y'}}
\end{equation}
where $\sigma_{\gamma,x}$ and $\sigma_{\gamma,y}$ are the \textit{rms} transverse sizes of the photon beam and $\sigma_{\gamma,x'}$ and $\sigma_{\gamma,y'}$ are the \textit{rms} transverse angular sizes of the photon beam. However, by taking advantage of the analogy to undulator radiation, it is possible to approximate the brilliance of the scattered photons using the parameters of the electron beam at the collision. The standard approximation is $\sigma_{\gamma,x'} \approx \sqrt{\epsilon_x/\beta_x + \lambda/2L}$, where $\epsilon_x$ and $\beta_x$ are parameters of the electron beam, $\lambda$ is the emitted wavelength, and $L$ is the effective length of the source. This result assumes the X-ray beam angular sizes are a combination of the intrinsic beam angles and radiation diffraction, which is quantified by $\lambda/2L$~\cite{RAST_Krafft}. However, the properties of compact sources are such that $\epsilon_{x,y} > \lambda/4\pi$, implying that the decrease in brilliance for the photon source due to $\lambda/2L$ terms is negligible~\cite{Handbook}. Taking these approximations into account, Eq.~(\ref{EQ:GEN_BRILLIANCE}) becomes
\begin{equation}
\mathcal{B} = \frac{\mathcal{F}_{0.1\%}}{4\pi^2\sigma_{\gamma,x}\sqrt{\epsilon_x/\beta_x}\sigma_{
\gamma,y}\sqrt{\epsilon_y/\beta_y}}.
\label{EQ:GEN_BRILLIANCE02}
\end{equation}

In previous papers~\cite{White_Paper, PRAB_16, Diss}, we have taken the approximation that the X-ray source size is the size of the electron beam; this approximation is typical in the characterization of compact sources~\cite{RAST_Krafft, NIMB_Jacquet}. In this approach, $\sigma_{\gamma,x} = \sigma_x = \sqrt{\beta_x\epsilon_x}$, so Eq.~(\ref{EQ:GEN_BRILLIANCE02}) becomes 
\begin{equation}
\begin{aligned}
\mathcal{B} &\approx \frac{\mathcal{F}_{0.1\%}}{4\pi^2\epsilon_x\epsilon_y} \\
 &\approx \frac{\gamma^2\mathcal{F}_{0.1\%}}{4\pi^2\epsilon^N_{x,\mathit{rms}}\epsilon^N_{y,\mathit{rms}}}.
\end{aligned}
\label{EQ:BRILLIANCE_OLD}
\end{equation}

If instead we take the position that the source size is a convolution of the electron and laser beam sizes, such that
\begin{equation}
\frac{1}{\sigma_{\gamma}^2} = \frac{1}{\sigma_{\mathit{laser}}^2} + \frac{1}{\sigma_x\sigma_y}.
\end{equation}
Using this, Eq.~(\ref{EQ:GEN_BRILLIANCE02}) becomes
\begin{equation}
\begin{aligned}
\mathcal{B} &\approx \frac{\mathcal{F}_{0.1\%}}{4\pi^2\sigma_{\gamma}^2\sqrt{\epsilon_x\epsilon_y/\beta_x\beta_y}} \\
 &\approx \frac{\gamma\mathcal{F}_{0.1\%}}{4\pi^2\sigma_{\gamma}^2\sqrt{\epsilon^N_{x,\mathit{rms}}\epsilon^N_{y,\mathit{rms}}/\beta_x\beta_y}}.
\end{aligned}
\label{EQ:BRILLIANCE_NEW}
\end{equation}
As the laser spot size becomes increasingly greater than the electron beam spot size, the difference between Eq.~(\ref{EQ:BRILLIANCE_OLD}) and Eq.~(\ref{EQ:BRILLIANCE_NEW}) becomes negligible. However, for the compact source presented in this paper, the spot sizes are roughly equivalent, making Eq.~(\ref{EQ:BRILLIANCE_NEW}) more appropriate. 

From either brilliance formula, it becomes clear that to maximize brilliance requires maximizing the photon flux or electron beam energy or minimizing the electron beam normalized \textit{rms} transverse emittances.

\section{\label{SEC:ICLS}Inverse Compton Light Source}

\subsection{\label{SEC:COMPTON_SCATTERING}Inverse Compton Scattering}

The process of scattering a photon off an electron at rest is known as Compton scattering. The term inverse Compton scattering (ICS) is used in the situation such that the electron loses energy to the incident photons. In the following formulae, $\Phi$ is the angle between the relativistic electron and the laser beams, and $\Delta\Theta$ is the angle between the laser beam and scattered photons. If $\theta$ and $\phi$ represent spherical polar angles that the scattered photons make in the coordinate system such that the electron beam moves along the $z$ axis, then the angle $\Delta\Theta$ is $\cos\Delta\Theta = \cos\Phi\cos\theta - \sin\Phi\sin\theta\cos\phi$. The coordinate system is set so the interaction point (IP) of the electron and laser beams occurs in the $x-z$ plane.

A general formula expressing the energy of a scattered photon in the lab frame, $E_{\gamma}$, as a function of the direction of the scattered photon, is
\begin{widetext}
\begin{equation}
E_{\gamma}(\Phi, \theta, \phi) = \frac{E_{\mathit{laser}}(1 - \beta\cos\Phi)}{1 - \beta\cos\theta + E_{\mathit{laser}}(1 - \cos\Phi\cos\theta + \sin\Phi\sin\theta\cos\phi)/E_{e^-}}
\end{equation}
\end{widetext}
where $\beta$ is the relativistic factor $v_z/c$, $E_{\mathit{laser}}$ is the energy of the typical laser photon, and $E_{e^-} = \gamma m_ec^2$ is the energy of the electron~\cite{RAST_Krafft}. This formula includes the impact of electron recoil. The Thomson formula is a good approximation if the electron recoil is negligible, i.e., the energy of the laser in the beam frame is much less than the rest mass of the electron. When this is true, then the formula for the energy of the scattered photon becomes
\begin{equation}
E_{\gamma}(\Phi, \theta) \approx E_{\mathit{laser}}\frac{1 - \beta\cos\Phi}{1 - \beta\cos\theta}.
\end{equation}
It can also be approximated as
\begin{equation}
E_{\gamma}(\Phi, \theta) \approx E_{\mathit{laser}}  \frac{2\gamma^2(1 - \beta\cos\Phi)}{1 + \gamma^2\theta^2},
\end{equation}
where $\gamma$ is the usual relativistic factor for the electron and $\gamma \gg 1$.

\begin{figure}
\includegraphics[width=\columnwidth]{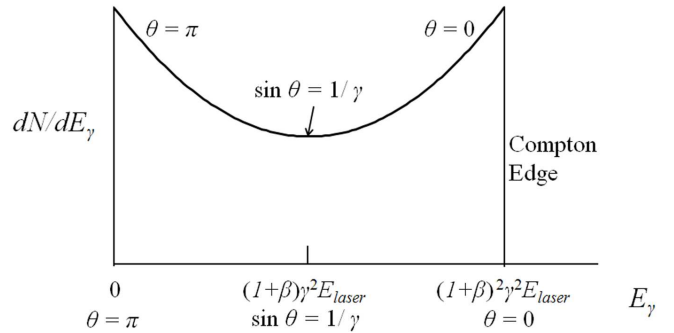}
\caption{Number density of scattered photons as a function of the energy of scattered photons. Annotated Fig.~2 from \cite{RAST_Krafft}.}
\label{FIG:COMPTON_EDGE}
\end{figure}

Consider the situation of a head-on collision between the electron and the laser ($\Phi = \pi$). The energy of the laser photon in the beam frame is $E'_{\mathit{laser}} = \gamma(1 + \beta)E_{\mathit{laser}}$. Assuming that the Thomson formula is a good approximation, i.e., $E'_{\mathit{laser}} \ll mc^2$ is true, then the energy of the scattered photon is also $E'_{\mathit{laser}}$ in the beam frame. Going back into the lab frame, the photons scattered in the forward (positive $z$) direction have the highest energy, which is $\gamma^2(1 + \beta)^2E_{\mathit{laser}} \approx 4\gamma^2E_{\mathit{laser}}$. The high energy boundary of emission is called the Compton edge; no radiation is emitted at higher energies. For photons scattered at the angle $\theta$ such that $\sin\theta = 1/\gamma$ $(1/\gamma \ll 1)$, the energy decreases to $2\gamma^2E_{\mathit{laser}}$, which is also the average energy of the scattered photons. Both the Compton edge and the number density of scattered photons as a function of the energy of scattered photons can be seen in Fig.~\ref{FIG:COMPTON_EDGE}.

The number of photons produced by scattering an incident laser off an electron is proportional to the time-integrated intensity of illumination. Consequently, the total photon yield is proportional to the square of the field strength, as in the case of undulator radiation. Progressing by the analogy with undulator radiation, the field strength parameter for a plane wave incident laser is defined to be
\begin{equation}
a = \frac{eE\lambda_{\mathit{laser}}}{2\pi mc^2},
\end{equation}
where $e$ is the electron charge, $E$ is the transverse electric field of the laser, $\lambda_{\mathit{laser}}$ is the laser wavelength, and $mc^2$ is the rest energy of the electron. This value represents the normalized transverse vector potential for the EM field accelerating the electrons during scattering. For Compton scattering, $a$ plays a role similar to that of $K$ in the field of undulators. For the case of $a \ll 1$, the backscattering is in the linear regime, an assumption that continues as formulae are presented.

If we take the assumption that the transverse intensity distributions of the laser and electron beams are round Gaussian distributions with the \textit{rms} sizes of $\sigma_e$ and $\sigma_{\mathit{laser}}$ respectively, then
\begin{equation}
U_{\gamma} = \gamma^2(1 + \beta)\sigma_T\frac{N_eN_{\mathit{laser}}}{2\pi(\sigma_e^2 + \sigma_{\mathit{laser}}^2)}E_{\mathit{laser}},
\end{equation}
where $U_{\gamma}$ is the total energy of the scattered photons per collision, $N_e$ is the number of electrons in the bunch, $N_{\mathit{laser}}$ the number of photons in the incident laser, and $\sigma_T$ is the Thomson cross section $8\pi r_e^2/3$, where $r_e$ is the classical electron radius~\cite{RAST_Krafft, Clarke_STUW, Duke_SRPP}. The classical electron radius is defined as $r_e = e^2/4\pi\epsilon_0mc^2$, where $e$ is the electric charge of the electron, $\epsilon_0$ is the permittivity of free space, $m$ is the mass of the electron, $c$ is the speed of light~\cite{Jackson}. From this formula, the total number of scattered photons $N_{\gamma}$ is
\begin{equation}
N_{\gamma} = \sigma_T\frac{N_eN_{\mathit{laser}}}{2\pi(\sigma_e^2 + \sigma_{\mathit{laser}}^2)}.
\end{equation}

Given that the spectral energy density of the scattered photons may be analytically computed in the linear Thomson backscatter limit, it can further be determined that the number of scattered photons within a $0.1\%$ bandwidth at the Compton edge is $N_{0.1\%} = 1.5\times10^{-3}N_{\gamma}$. Consequently, the rate of photons (flux) into this bandwidth is $\mathcal{F}_{0.1\%} = 1.5\times10^{-3}\dot{N_{\gamma}}$. For high-frequency repetitive sources, $\dot{N_{\gamma}} = fN_{\gamma}$, where $f$ is the repetition rate~\cite{RAST_Krafft, Clarke_STUW, Duke_SRPP}.

\subsection{Compact Sources}

There are two main components in an inverse Compton light source (ICLS) - a relativistic electron beam and an incident laser. In the last several years, there has been a significant advancement in the technology to produce a suitable laser. The details of this progress are largely beyond the scope of this article, though the status of the current technology will be touched on later. The other component, the focus of this project, is the relativistic electron beam off which the incident laser scatters.

There exist two schemes for accelerating an electron beam to the desired energy, typically in the range of a few $10$s of MeV: a linear accelerator (linac) or a storage ring (ring)~\cite{NIMB_Jacquet}. A linac is composed of radiofrequency (RF) or superconducting (SC) RF (SRF) cavities that accelerate the beam to the desired energy~\cite{Wiedemann}. Rings are circular devices into which a beam of a specific energy is injected, where the beam may or may not be extracted before being used~\cite{Handbook}.

Both of these options have benefits and drawbacks. Existing storage ring projects typically have lower expected fluxes than those of linacs. The expected brightness is frequently lower~\cite{NIMB_Jacquet}, as the smallest achievable normalized emittances are typically larger for a ring than a linac. Additionally, a full energy linac is often required anyway for injection into the ring~\cite{Wiedemann, NIMB_Jacquet, Newer_Graves}. However, rings are capable of a high repetition rate, a higher average current than is typical for linacs, and historically have better stability~\cite{NIMB_Jacquet, Newer_Graves}.

Linac-based ICS X-ray sources have shown promising results at lower pulse repetition rates, though these results have yet to be reproduced at higher rates. For electron beams with an energy above 10~MeV, cumbersome shielding must be included~\cite{Newer_Graves, NIMB_Jacquet}. Current cryogenic equipment for SRF structures, which are utilized in all but one of the known linac projects (and, indeed, are by some assumed to be necessary for a linac project to succeed), are more complicated than non-expert users are comfortable using. Another common feature to most linac projects is a superconducting electron gun, a technology with promising results but not yet a mature field~\cite{Arnold_Guns, NIMB_Jacquet}. Linac projects are more likely to be capable of shorter bunch lengths, even without compression, smaller normalized emittances, and a greater flexibility for phase space manipulations than ring projects~\cite{NIMB_Jacquet, Newer_Graves}.

Referenced in the literature as the first existing compact Compton source is the one built by Lyncean Technologies. An electron beam is produced by a normal conducting linac and injected into a storage ring, which occupies a 1 m by 2 m footprint. This machine delivers ${\sim}10^9$ ph/s in a $3 \%$ energy bandwidth, with the scattered photon beam having an \textit{rms} spot size of ${\sim}$45~$\mu$m~\cite{NIMB_Jacquet, Lyncean}.

Table~\ref{TB:XRAY_COMP} contains some of the current projects with a compact ICLS design. To give some perspective to these values, the best rotating anodes, such as may currently be found in a lab as an X-ray source, have a flux of ${\sim}6 \times 10^9$ ph/s and a brightness on the order of $10^9$ photons/(sec-mm\textsuperscript{2}-mrad\textsuperscript{2}-0.1\%BW)~\cite{Newer_Graves}. On the other hand, an X-ray beam that might typically be found at a large facility has a flux in the regime of ${\sim}10^{11} - 10^{13}$ ph/s~\cite{APS_Beamlines} and a brightness of ${\sim}10^{19}$ photons/(sec-mm\textsuperscript{2}-mrad\textsuperscript{2}-0.1\%BW)~\cite{NIMA_APS_figs}.

Given these numbers, a robust user program for a compact ICLS machine would require that substantial fluxes of narrow-band X-rays are the desired requirement, rather than the best average brightness. However, the potential for such machines, in terms of both performance and demand, make the prospect of a well-designed compact source non-negligible~\cite{RAST_Krafft}.

\begin{table*}
\caption{Comparison of X-ray beam parameters for different ICLS compact designs.}
\label{TB:XRAY_COMP}
\begin{ruledtabular}
\begin{tabular}{lccccc}
Project & Type & $E_x$ (keV) & Ph/s & Ph/(s-mrad\textsuperscript{2} & $\sigma_{\gamma}$ ($\mu$m) \\
 & & & & -mm\textsuperscript{2}-0.1\%BW) & \\
\hline
Lyncean at Munich~\cite{MunichLightSource, Lyncean, RAST_Krafft} & SR & 10-20 & $10^{11}$ & $10^{11}$ & 45 \\
TTX~\cite{Jacquet_10} & SR & 20-80 & $10^{12}$ & $10^{10}$ & 50 \\
LEXG~\cite{Jacquet_11} & SR (SC) & 33 & $10^{13}$ & $10^{11}$ & 20 \\
ThomX~\cite{Jacquet_13} & SR & 20-90 & $10^{13}$ & $10^{11}$ & 70 \\
KEK QB~\cite{Jacquet_14} & Linac (SC) & 35 & $10^{13}$ & $10^{11}$ & 10 \\
KEK ERL~\cite{Jacquet_15} & Linac (SC) & 67 & $10^{13}$ & $10^{11}$ & 30 \\
NESTOR~\cite{Jacquet_12} & SR & 30-500 & $10^{13}$ & $10^{12}$ & 70 \\
ASU (MIT)~\cite{Newer_Graves} & Linac & 12 & $10^{13}$ & $10^{12}$ & 2 \\
\textbf{ODU (12 microns)} & \textbf{Linac (SC)} & \textbf{12} & $\mathbf{10^{13}}$ & $\mathbf{10^{14}}$ & \textbf{3} \\
\textbf{ODU (3.2 microns)} & \textbf{Linac (SC)} & \textbf{12} & $\mathbf{10^{14}}$ & $\mathbf{10^{15}}$ & \textbf{2} \\
\end{tabular}
\end{ruledtabular}
\end{table*}

\subsection{Considerations for This Project and Design Parameters Choices}

The main goal of this study was to develop the concept for a high-brilliance, high-flux inverse Compton light source that would also be relatively affordable and easy to operate by  non-experts.  High-flux would imply cw operation and an SRF linac, and ease of operation would suggest operating with atmospheric helium at 4.2 K or above.  Since the surface resistance of superconductors increases quadratically with frequency this would imply a low-frequency system.  On the other hand, the size and cost of the cavities and cryomodules increase as the frequency is lowered,  and a trade-off between the two considerations suggested a frequency range of 300 to 500 MHz~\cite{White_Paper}.

A number of accelerating structure geometries were considered. The most common and widespread is the TM-type, sometimes referred to as ``elliptical''. This geometry is well-understood and has the advantage of having rotational symmetry.  However it was deemed to be too large in that frequency range. Another type of superconducting structure is the spoke geometry which, at the same frequency, is smaller than the TM-type \cite{DoubleSpokeUsage_04, DoubleSpokeUsage_06}. Several of these cavities have been developed in the frequency and velocity range of interest \cite{SpokeTest_02}. A 325 MHz single-spoke cavity had been successfully developed but was also deemed to be too large \cite{SpokeTest_01}. We finally decided on a 500 MHz, double-spoke geometry which had also been successfully developed \cite{SpokeTest_03}. The spoke geometry has the disadvantage of introducing quadrupole components in the electromagnetic fields \cite{NAPAC_13}. As shown later, the contribution of the quadrupole components can be managed and its impact on the final performance is minimal.

We would like to emphasize that further advances in the SRF technology could justify revisiting the geometry and/or frequency choice but would not invalidate the conclusions of this study.

To increase brightness, the normalized \textit{rms} transverse emittance needs to be minimized, leading to a target value of 0.1~mm-mrad. While this value is considerably smaller than in other SRF injector guns, as shown in this work, a low bunch charge of 10~pC makes this emittance attainable~\cite{White_Paper, Arnold_Guns}. To attain a high average flux, considering that the average flux is proportional to both the bunch charge and the repetition rate, a high repetition rate of 100~MHz was chosen to counterbalance the low bunch charge. Minimizing the spot size of both electron and laser beams also helped to increase the flux. Thus, the spot size for the electron beam at the IP was set at ${\sim}$3 $\mu$m, which is small but feasible, though it will require state-of-the-art diagnostics at the IP. 

An electron beam energy of 25~MeV and an incident scattering laser energy of 1.24~eV were chosen. The chosen energies generate X-rays of up to 12~keV. X-rays at 12~keV have a corresponding wavelength of approximately one Angstrom, the same as in large third generation synchrotron facilities. For the energy smearing of the forward flux to be small relative to the total bandwidth necessitates that the relative beam energy spread be less than 0.03\%. At the chosen energy of 25~MeV, this leads to an \textit{rms} energy spread requirement of 7.5~keV. In order to keep the flux reduction due to the hourglass effect negligible, the compressed bunch length is set to less than 1~mm~\cite{White_Paper}.

For the best possible X-ray beam, a high quality high power laser is necessary. The ideal laser would, among other properties, have a circulating power of 1~MW, compared to 100~kW today. One MW is widely regarded as feasible, but has not yet been achieved in a compact optical cavity~\cite{White_Paper, RAST_Krafft, MIT_Paper, NIMB_Jacquet}. The other properties relevant to the optical cavity are less demanding: 1~$\mu$m wavelength (1.24~eV), $5 \times 10^{16}$ ph/bunch, spot size of 3.2~$\mu$m at collision, and peak strength parameter $a = 0.026$~\cite{White_Paper}. However, a 3.2~$\mu$m laser spot size has an extremely short Rayleigh range (which presents additional challenges), so results are also presented for a laser spot size of 12~$\mu$m. 

It is possible to take the properties of the electron beam and incident laser beam and estimate selected parameters of the X-ray beam which would be produced from a collision between the two, using formulae presented previously. For a laser spot size of 3.2~$\mu$m, the X-ray beam energy will be 12~keV with $1.6 \times 10^6$ photons/bunch. The X-ray beam flux will be $1.6 \times 10^{14}$ ph/s, with an average brilliance of $3 \times 10^{15}$ ph/(sec-mm\textsuperscript{2}-mrad\textsuperscript{2}-0.1\%BW). For a laser spot size of 12~$\mu$m, the X-ray beam energy will be 12~keV with $2.1 \times 10^5$ photons/bunch. The X-ray beam flux will be $2.1 \times 10^{13}$ ph/s, with an average brilliance of $2.1 \times 10^{14}$ ph/(sec-mm\textsuperscript{2}-mrad\textsuperscript{2}-0.1\%BW). These values are sufficiently high as to indicate that a compact Compton source which fulfills these parameters is likely to be very interesting to potential users~\cite{NIMB_Jacquet}.

These specifications are based on and similar to those earlier presented in \cite{MIT_Paper}. Desired electron beam parameters at the interaction point (IP) are shown in Table~\ref{TB:BEAM_IP_DESIRED}. Optical cavity parameters are shown in Table~\ref{TB:LASER_IP_DESIRED}, based on performances that may soon be attainable~\cite{MIT_Paper, RAST_Krafft}. Using the values in these tables and the formulae previously presented, the resulting X-ray beam can be described by the quantities in Table~\ref{TB:XRAY_DESIRED} for the proposed laser spot sizes. 

\begin{figure*}
\includegraphics[width=0.39\textwidth]{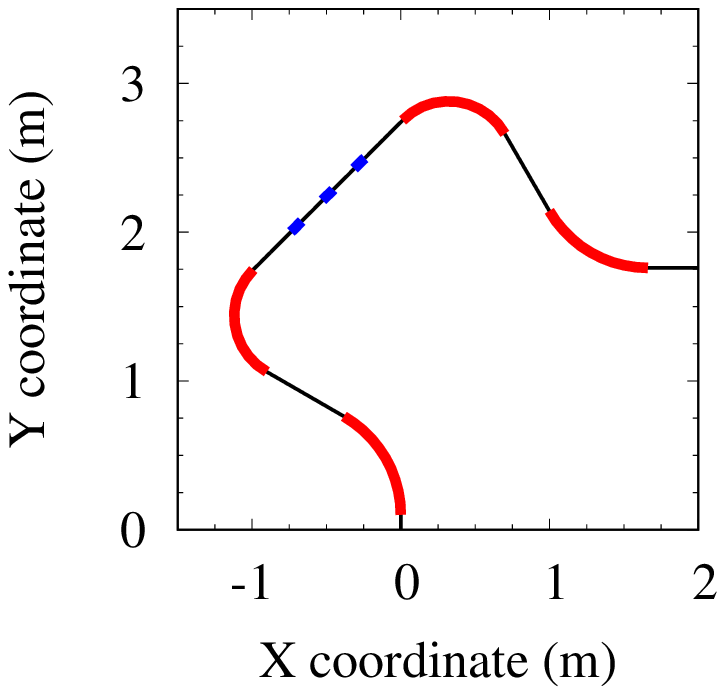}
\includegraphics[width=0.59\textwidth]{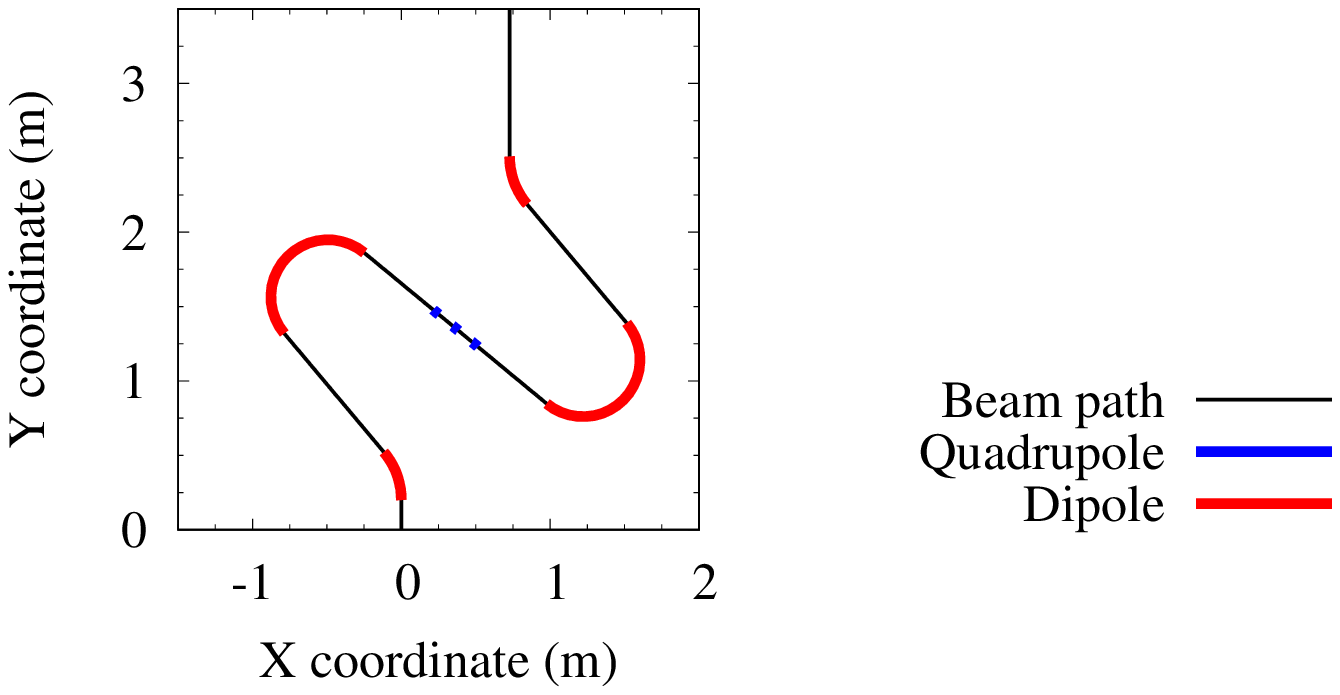}
\caption{Basic layout examples of $3\pi$ (left) and $4\pi$ (right) bunch compressors. Beam enters at (0, 0). }
\label{FIG:COMPRESSORS}
\end{figure*}

One of the benefits of this design is that the layout is entire linear, which can be seen in Fig.~\ref{FIG:ENTIRE}. This benefit allows for a simpler and more compact design. While we have seen improvement in the transverse emittance by increasing the length of the bunch off the cathode, a longer bunch length requires a bunch compressor, increasing both the size and complexity of the design. The bunch compressor might be a $3\pi$ or $4\pi$ design, basic examples of which are shown in Fig.~\ref{FIG:COMPRESSORS}\cite{Diss,Randy_NAPAC_13}.

\begin{table}
\caption{Desired electron beam parameters at interaction point.}
\label{TB:BEAM_IP_DESIRED}
\begin{ruledtabular}
\begin{tabular}{lcc}
Parameter & Quantity & Units \\
\hline
Energy & 25 & MeV \\
Bunch charge & 10 & pC \\
Repetition rate & 100 & MHz \\
Average current & 1 & mA \\
Transverse \textit{rms} normalized emittance & 0.1 & mm-mrad \\
$\beta_{x,y}$ & 5 & mm \\
$\sigma_{x,y}$ & 3 & $\mu$m \\
FWHM bunch length & 3 (0.9) & psec (mm) \\
\textit{rms} energy spread & 7.5 & keV \\
\end{tabular}
\end{ruledtabular}
\end{table}

\begin{table}
\caption{Laser parameters at interaction point.}
\label{TB:LASER_IP_DESIRED}
\begin{ruledtabular}
\begin{tabular}{lcc}
Parameter & Quantity & Units \\
\hline
Wavelength & 1 (1.24) & $\mu$m (eV) \\
Circulating power & 1 & MW \\
$N_{\gamma}$, Number of photons/bunch & $5 \times 10^{16}$ & \\
Spot size (\textit{rms}) & 3.2, 12 & $\mu$m \\
Peak strength parameter, $a$ & 0.026, 0.002 & \\
\hspace{5mm}$a = eE\lambda_{\mathit{laser}}/2\pi mc^2$ & & \\
Repetition rate & 100 & MHz \\
\textit{rms} pulse duration & 2/3 & ps \\
\end{tabular}
\end{ruledtabular}
\end{table}

\begin{table}
\caption{Desired light source parameters.}
\label{TB:XRAY_DESIRED}
\begin{ruledtabular}
\begin{tabular}{lccc}
Parameter & \multicolumn{2}{c}{Laser spot ($\mu$m)} & Units \\
 & 3.2 & 12 & \\
\hline
X-ray energy & Up to 12 & Up to 12 & keV \\
Photons/bunch & $1.6 \times 10^6$ & $2.1 \times 10^5$ &  \\
Flux & $1.6 \times 10^{14}$ & $2.1 \times 10^{13}$ & ph/sec \\
Average brilliance & $3.0 \times 10^{15}$ & $2.1 \times 10^{14}$ & ph/(s-mm\textsuperscript{2}\\
 & & & -mrad\textsuperscript{2}-0.1\%BW) \\
\end{tabular}
\end{ruledtabular}
\end{table}

\section{\label{SEC:GUN}SRF Gun}

\subsection{Similar Design Comparison}

There exist three types of photoinjectors, or guns, presently: the DC gun, the normal conducting RF gun, and the SRF gun. While the first two types represent technology that is mature and the result of development over many decades, SRF guns are still an emerging technology~\cite{Arnold_Guns, Newer_Graves}.

The concept for an SRF gun was initially published in the early 1990s~\cite{Michalke_Dissertation}, though more consistent publishing on the subject did not occur until nearly a decade later~\cite{Michalke_SRF, Michalke_EPAC, Pagani, Janssen_First}. Using the idea of a reentrant cavity for an SRF gun was first presented in \cite{Michalke_SRF}, which subsequently inspired a number of similar gun designs~\cite{Harris, Arnold_Guns}. Table~\ref{TB:GUN_COMP} compares various SRF gun designs with each other and to the parameters ultimately achieved by this study, referred to as ODU ICLS in the table. This table contains the design parameters for projects at the Naval Postgraduate School (NPS), the University of Wisconsin FEL (WiFEL), and Brookhaven National Lab (BNL).

\begin{table}
\caption{Comparison of various SRF gun design projects~\cite{Harris,Arnold_Guns}.}
\label{TB:GUN_COMP}
\begin{ruledtabular}
\begin{tabular}{lccccc}
Parameter & ODU ICLS & NPS & WiFEL & BNL & Units \\
\hline
Frequency & 500 & 500 & 200 & 112 & MHz \\
Bunch charge & 0.01 & 1 & 0.2 & 5 & nC \\
Trans. norm. & 0.1, 0.13 & 4 & 0.9 & 3 & mm-mrad\\
\textit{rms} emittance  & & & & & \\
\end{tabular}
\end{ruledtabular}
\end{table}

There are two considerations that can be seen from Table~\ref{TB:GUN_COMP}. The first is that the bunch charge of the ODU ICLS gun is smaller than the other designs by an order of magnitude or more. The second is that the desired transverse normalized \textit{rms} emittance is also smaller than the other designs by nearly an order of magnitude or more. This reduced bunch charge is what makes the extremely small emittance feasible.

It is common in RF/SRF gun design to mitigate the growth of the transverse emittance of the bunch due to space charge in order to produce a beam with the smallest emittance. Emittance compensation is the reduction of emittance due to linear space-charge forces~\cite{Carlsten,Serafini}. One of the most common techniques in emittance compensation is the use of a solenoid. By placing a solenoid after an injector, the goal is to manipulate the transverse phase space so that the focusing effect of the solenoid negates the defocusing effect of the space charge~\cite{Serafini, Carlsten, Kumar}. This technique is used in the three other SRF gun designs listed in Table~\ref{TB:GUN_COMP}~\cite{Arnold_Guns}.

At the beginning of this study, simulations were run that modeled a bunch exiting the gun which passed through a solenoid before entering the linac. This approach to emittance compensation failed in two ways - the transverse normalized \textit{rms} emittance was not decreased and the bunch exiting the linac was difficult to manipulate for compression and final focusing~\cite{White_Paper}. Consequently, in designing the ODU ICLS accelerator a different approach was taken, which utilized RF focusing by altering gun geometry to provide focusing, instead of it being provided by a solenoid as in similar SRF gun designs~\cite{Arnold_Guns}.

\begin{figure*}
\includegraphics[width=0.95\columnwidth]{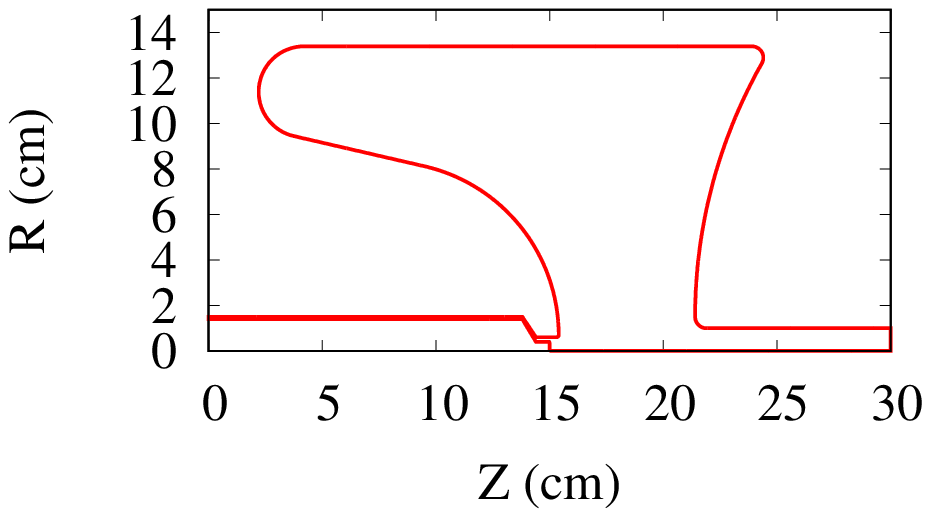}
\includegraphics[width=0.95\columnwidth]{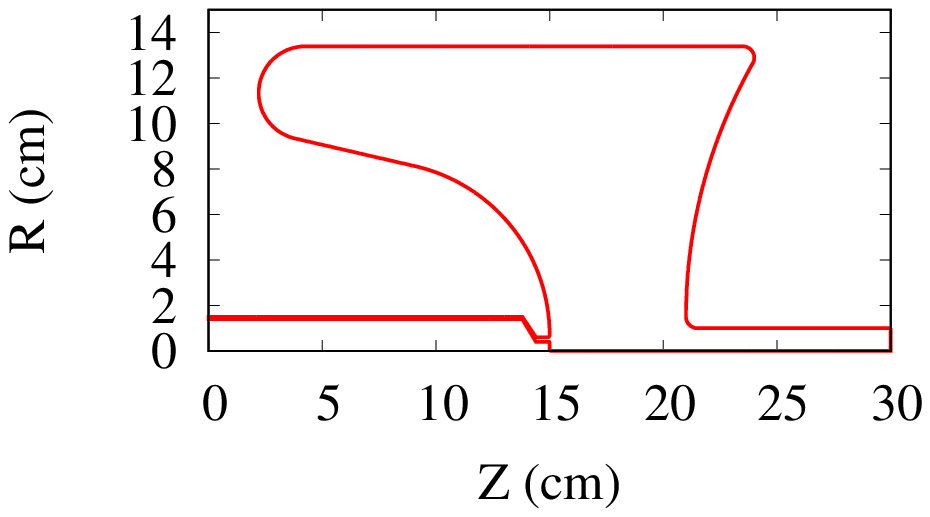}
\caption{Two identical gun geometries with (left) and without (right) a recessed cathode to provide RF focusing.}
\label{FIG:RF_FOCUSING}
\end{figure*}

RF focusing refers to focusing provided by the RF EM fields of the accelerating structure~\cite{Wangler}. One example of this is shown in \cite{RF_Focusing}, where the RF EM fields of the gun are manipulated by recessing the cathode holder by a varying amount. In Fig.~\ref{FIG:RF_FOCUSING}, two similar gun geometries are shown, with the only difference between them being the recessed cathode in the bottom right figure. In essence, this alteration to the gun geometry is to produce a radial electric field which focuses the beam. Ideally, the focusing provided will negate the defocusing produced by the space charge. However, there is a cost to this approach. As the cathode is further recessed, the radial component of the electric field (and thus the focusing) increases, but the longitudinal component (which accelerates the beam) decreases~\cite{RF_Focusing}.

By changing the geometry of the nosecone, it is also possible to alter the EM fields within the gun. Regardless of how the radial field is produced, there is still a balancing act that must be found between the accelerating and focusing fields. Given that increasing the focusing field decreases the accelerating one, a simplistic line of thought leads one to simply increase the operating gradient until the bunch that exits the gun is sufficiently relativistic such that space charge is negligible. There are two main reasons that such an approach is not feasible.

First, for any given gun geometry there is a point at which increasing the operating gradient is more detrimental than beneficial to the beam quality. Past this point, the strength of the focusing field is actually over-compensating for the effects of space charge on the bunch, increasing the emittance at the exit of the linac. Therefore, in general there exists an operating gradient for a given geometry which produces the smallest transverse emittance, analogous to choosing the correct lens focal length to focus a beam of light at a particular location. Second, there exists a maximum threshold for surface fields on an SRF structure for reliable function. As the operating gradient is directly proportional to the surface fields, a maximum threshold for the operating gradient exists for any given geometry~\cite{Diss}.

\subsection{Initial Bunch Distribution and Drive Laser}

The initial bunch distribution off the cathode has the properties given in Table~\ref{TB:BUNCH_03}. This bunch is long enough to make longitudinal space charge effects negligible, while short enough to remove the need for a bunch compressor, which simplifies the design~\cite{Diss}. In order to produce a 4.5~psec flat-top bunch off the cathode, there exist multiple options. One fully realized option is in use in the LCLS injector~\cite{Akre_LCLS}. This drive laser was manufactured by \textit{Thales Laser} and is a frequency tripled, chirped-pulse amplification system based on a Ti:sapphire laser~\cite{Akre_3,Akre_LCLS}. The specifications called for by the LCLS commissioning require a FWHM pulse duration of 6~ps with a repetition rate of up to 120~Hz. In addition, the laser has an adjustable pulse duration between 3 and 20~ps~\cite{Akre_LCLS}. While the pulse duration is in the correct regime this project requires, the repetition rate is less than required by nearly two orders of magnitude.

Another scheme for producing a flat-top bunch off the cathode involves the use of long-period fiber gratings (LPGs). Using this approach, it has been demonstrated experimentally that Gaussian-like optical pulses can be transformed into flat-top pulses. In the proof of concept experiment which confirmed this approach, 600~fs and 1.8~ps Gaussian-like pulses were transformed into 1 and 3.2~ps flat-top pulses, respectively. The same LPG was used for both transformations, demonstrating the adaptability of such a device~\cite{Park_Laser_Solution}. It remains to demonstrate this technology at high average power.

\begin{table}
\caption{Bunch distribution off the cathode.}
\label{TB:BUNCH_03}
\begin{ruledtabular}
\begin{tabular}{lcc}
Parameter & Quantity & Units \\
\hline
Longitudinal distribution & Plateau & \\
Bunch length & 4.5 & ps \\
Rise time & 1.125 & ps \\
Radial distribution & Uniform & \\
\textit{rms} bunch radius & 1 & mm \\
Initial transverse momentum & 0 & mrad \\
Bunch charge & 10 & pC \\
Initial kinetic energy & 1 & keV \\
$p_z$ distribution & Isotropic & \\
\end{tabular}
\end{ruledtabular}
\end{table}

\subsection{Optimization Leading to the Geometry}

\begin{figure*}
\includegraphics[width=\textwidth]{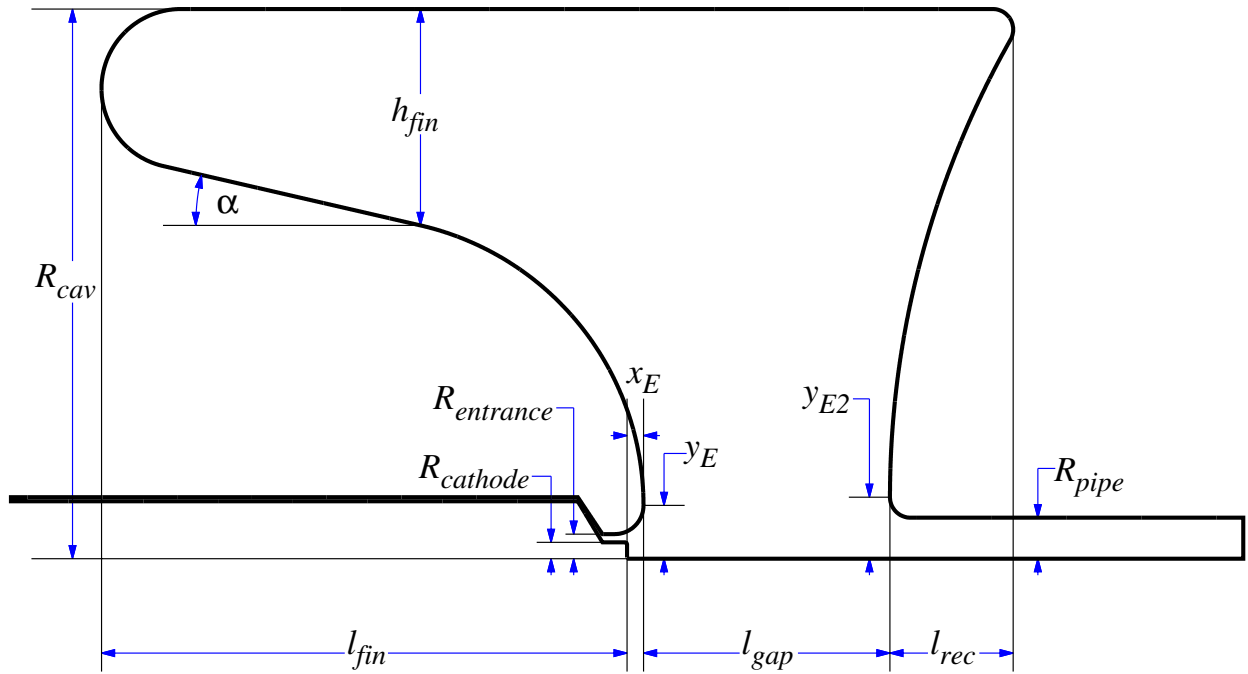}
\caption{Diagram of gun geometry with parameters.}
\label{FIG:GUN_LABELS}
\end{figure*}

During the course of the design further gun optimization was necessary to obtain the desired electron beam at the IP. To support the optimization it was necessary to create a set of parameters to fully define the parametric piecewise function that describes the gun shape, assuming the overall gun shape is retained. A set of formulae was created that required twelve parameters, shown in Fig.~\ref{FIG:GUN_LABELS}.

While a cursory examination was made of different parameters, $y_E$ is the key parameter to change to produce a suitable electron beam at the interaction point. This parameter being key is not surprising, given its proximity to both the center of the gun and the cathode holder. By altering this parameter over a range of values and evaluating the electron beam at the exit of the linac, the gun geometry was chosen. Further optimization of the other parameters may produce a better design at a later date.

\subsection{Final Geometry and Simulation Results}

\begin{figure}
\includegraphics[width=0.7\columnwidth]{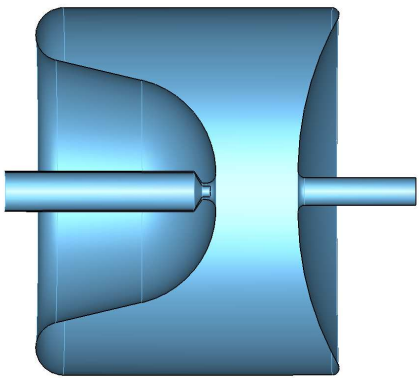}
\caption{SRF gun geometry.}
\label{FIG:GUN}
\end{figure}

\begin{figure*}
\includegraphics[width=0.95\textwidth]{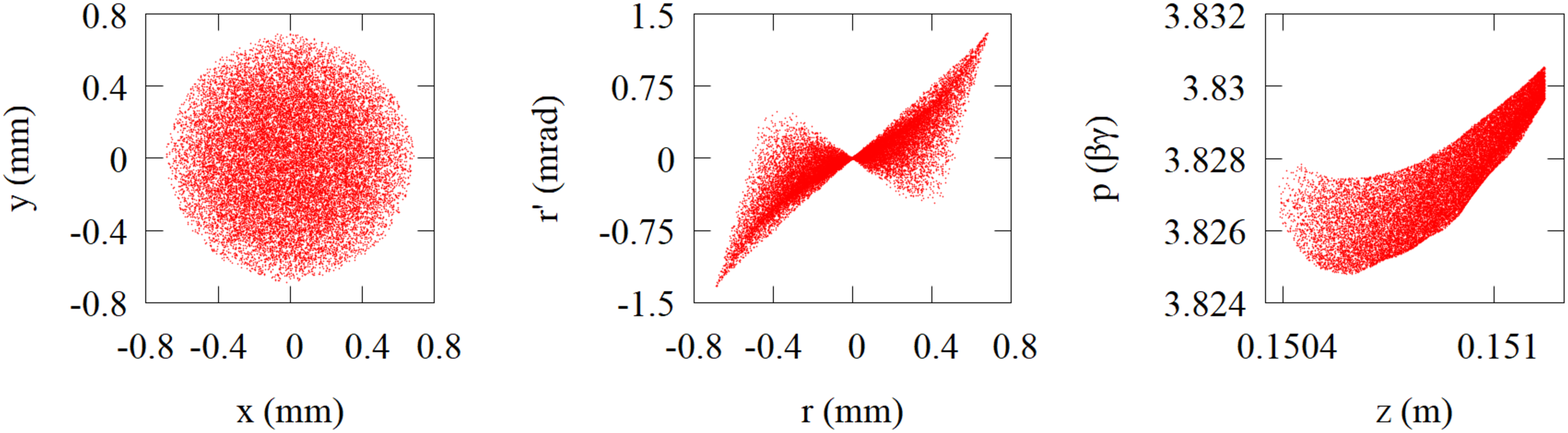}
\caption{Beam spot (left), transverse phase space (center), and longitudinal phase space (left) of bunch exiting gun.}
\label{FIG:FINAL_GUN}
\end{figure*}

The optimized geometry is shown in Fig.~\ref{FIG:GUN}, with the physical and RF properties given in Table~\ref{TB:GUN_SRF}. We used IMPACT-T~\cite{IMPACTT} to track 100,000 macroparticles through the EM fields simulated by Superfish~\cite{Superfish}. The tracking results at the exit of the gun are shown in Table~\ref{TB:FINAL_GUN_TRACKING}, with the transverse phase space and beam spot at the gun exit shown in Fig.~\ref{FIG:FINAL_GUN}.

\begin{table}
\caption{Cavity and RF properties of the gun design. Set to operate at $E_{\mathrm{acc}} = 10.3$~MV/m.}
\label{TB:GUN_SRF}
\begin{ruledtabular}
\begin{tabular}{lcc}
Parameter & Quantity & Units \\
\hline
Frequency of accelerating mode & 500 & MHz \\
Cavity length & 221.5 & mm \\
Cavity radius & 134 & mm \\
Cavity gap & 69 & mm \\
Beamport aperture radius & 10 & mm \\
\hline
Peak electric surface field $E_p^*$ & 3.86 & MV/m \\
Peak magnetic surface field $B_p^*$ & 6.55 & mT \\
$B_p^*/E_p^*$ & 1.70 & mT/(MV/m) \\
Geometrical factor, $G$ & 83.7 & $\Omega$ \\
$(R/Q) \times G$ & $1.31 \times 10^4$ & $\Omega^2$ \\
Energy content $U^*$ & 44 & mJ \\
\hline
\multicolumn{3}{l}{$^*$At $E_{\mathrm{acc}} = 1$ MV/m} \\
\end{tabular}
\end{ruledtabular}
\end{table}

\begin{table}
\caption{IMPACT-T tracking results at gun exit.}
\label{TB:FINAL_GUN_TRACKING}
\begin{ruledtabular}
\begin{tabular}{lcc}
Parameter & Quantity & Units \\
\hline
kinetic energy & 1.51 & MeV \\
\textit{rms} energy spread & 0.68 & keV \\
$\sigma_{x,y}$ & 0.29 & mm \\
$\epsilon^N_{(x,y), \mathrm{rms}}$ & 0.20 & mm-mrad \\
$\sigma_z$ & 0.18 & mm \\
\end{tabular}
\end{ruledtabular}
\end{table}

\section{\label{SEC:LINAC}Linac}

\subsection{SRF Double-Spoke Cavity}

Until recently, accelerating electrons near the speed of light has not been attempted with multi-spoke cavities. This is largely because of the well-established and successful performance of TM-type cavities. However, multi-spoke cavities are familiar options for accelerating ions. Previous studies of multi-spoke cavities for $\beta \sim$ 1 strongly suggest that they are a viable option for accelerating electrons~\cite{DoubleSpokeUsage_01, DoubleSpokeUsage_02, DoubleSpokeUsage_03, DoubleSpokeUsage_04, DoubleSpokeUsage_05}.

\begin{figure}
\includegraphics[width=\columnwidth]{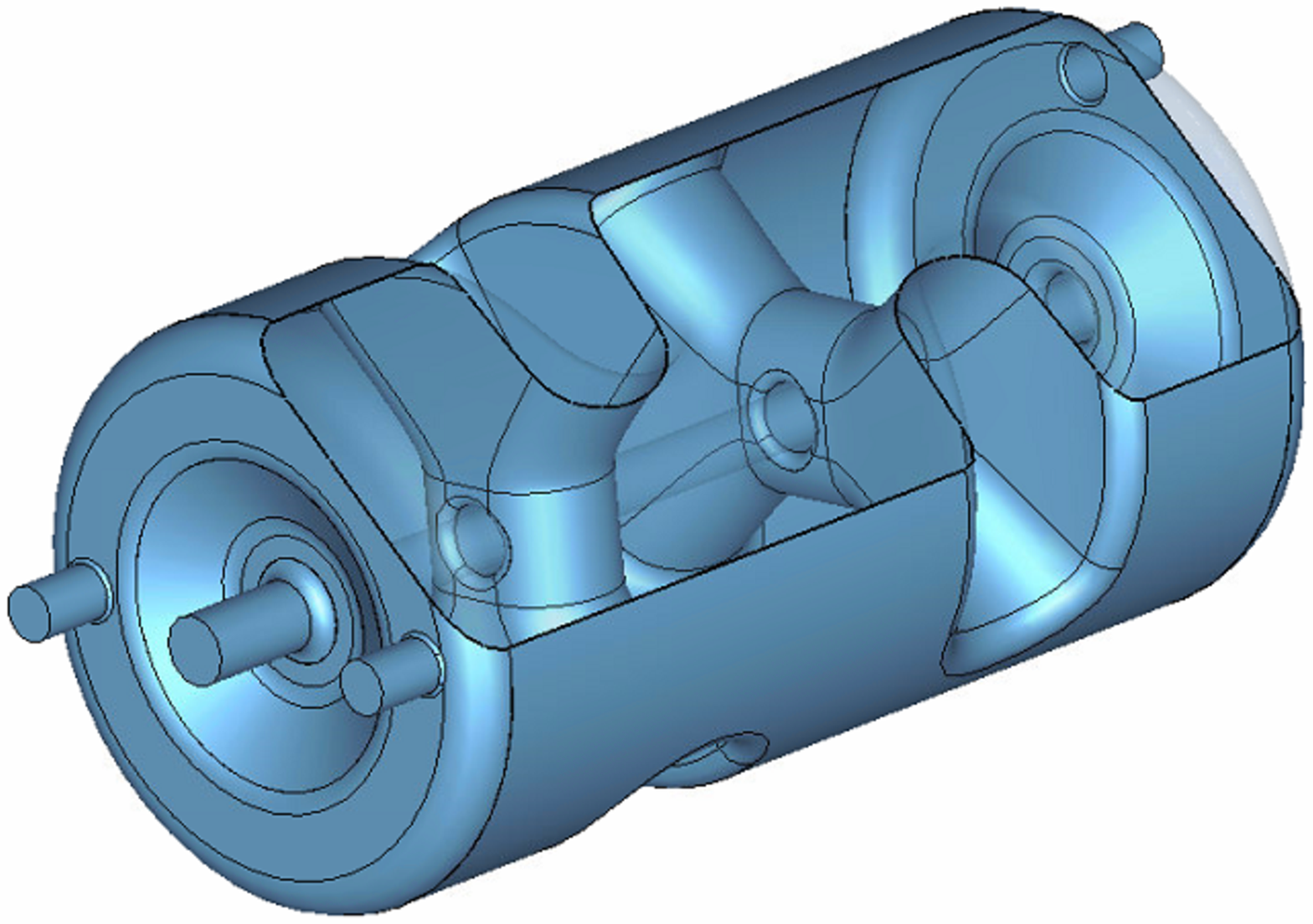}
\caption{The double-spoke SRF cavity, with a portion cut away to display the interior structure.}
\label{FIG:DOUBLESPOKE}
\end{figure}

\begin{figure}
\includegraphics[width=\columnwidth]{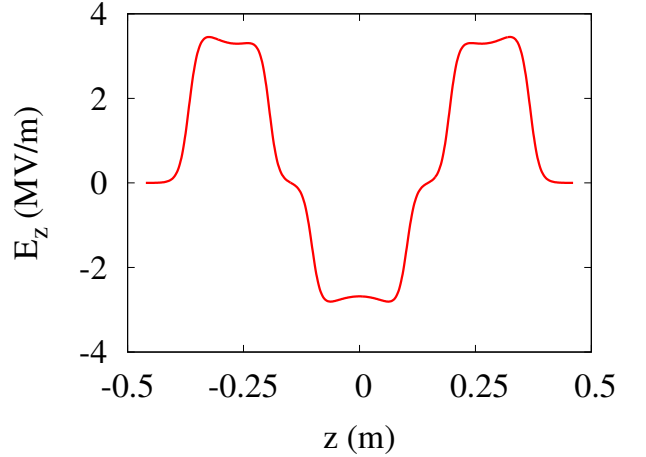}
\caption{The accelerating electric field along the beamline of the double-spoke SRF cavity.}
\label{FIG:DOUBLESPOKE_FIELD}
\end{figure}

The four 500~MHz cavities which comprise the linac are double-spoke speed-of-light SRF cavities designed by Christopher Hopper during his ODU PhD research~\cite{Chris_Thesis, Chris_PRAB, NAPAC_13}. The electron beam gains nearly 5.9~MeV as it passes through each cavity in the linac. Fig.~\ref{FIG:DOUBLESPOKE} contains an image of this cavity, with a portion cut away to more clearly view the interior structure. The accelerating field of this cavity is shown in Fig.~\ref{FIG:DOUBLESPOKE_FIELD}, with the complete EM field calculated by CST MICROWAVE STUDIO\textsuperscript{\textregistered}~\cite{CST_MWS}. Select RF and physical properties are contained in Table~\ref{TB:DOUBLESPOKE}. For more information on the optimization of the double-spoke cavity design, the reader is directed toward \cite{Chris_PRAB}. It is difficult to shorten the linac without requiring gradients which may not be reliably achieved.

\begin{table}
\caption{Physical (top) and RF (bottom) properties of double-spoke cavity.}
\label{TB:DOUBLESPOKE}
\begin{ruledtabular}
\begin{tabular}{lcc}
Parameter & Quantity & Units \\
\hline
Frequency of accelerating mode & 500 & MHz \\
Frequency of nearest mode & 507.1 & MHz \\
Cavity diameter & 416.4 & mm \\
Iris-to-iris length & 725 & mm \\
Cavity length & 805 & mm \\
Reference length [(3/2)$\beta_0\lambda$] & 900 & mm \\
Aperture diameter & 50 & mm \\
\hline
Energy gain$^*$ at $\beta_0$ & 900 & kV \\
$R/Q$ & 675 & $\Omega$ \\
$QR_s^{\dag}$ & 174 & $\Omega$ \\
$(R/Q)\times QR_s^{\dag}$ & 1.2$\times10^{5}$ & $\Omega^2$ \\
Peak electric surface field $E_p^*$ & 3.7 & MV/m\\
Peak magnetic surface field $B_p^*$ & 7.6 & mT \\
$B_p^*/E_p^*$ & 2.05 & mT/(MV/m) \\
Energy content$^*$ & 0.38 & J \\
Power dissipation$^{*\dag}$ & 0.87 & W \\
\hline
\multicolumn{3}{l}{$^*$At $E_{\mathrm{acc}} = 1$ MV/m and reference length $(3/2)\beta_0\lambda$, $\beta_0$ = 1}\\
\multicolumn{3}{l}{$^{\dag}R_s$ = 125 n$\Omega$} \\
\end{tabular}
\end{ruledtabular}
\end{table}

\subsection{Layout and Simulation Results}

One aspect of the double-spoke cavity is the ``quadrupole-like'' behavior of the cavities - the electron beam is focused in $x$ and defocused in $y$, or vice versa by the accelerating mode~\cite{Wiedemann, LINAC2012_multipoles, NAPAC_13}. This aspect means that some adjustment is necessary to provide a round beam spot to the bunch compressor or final focusing section. When arranging the double-spoke cavities, the center two cavities are rotated 180$^{\circ}$ around the $y$-axis, as seen in Fig.~\ref{FIG:ENTIRE}. Simulations have demonstrated that this layout produces the roundest beam at the exit of the linac.

\begin{figure*}
\includegraphics[width=0.7\textwidth]{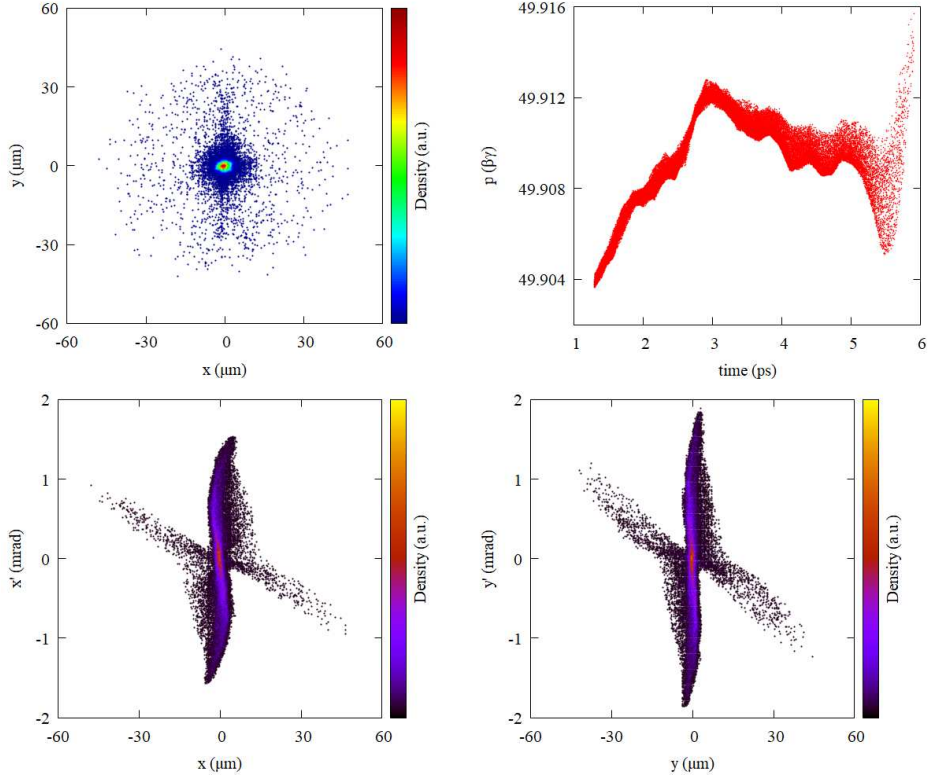}
\caption{Beam spot (upper left), longitudinal phase space (upper right), horizontal phase space (bottom left), and vertical phase space (bottom right) of bunch after exiting the linac.}
\label{FIG:LINAC}
\end{figure*}

Continuing to simulate the beam past the gun exit yields the electron beam properties given in Table~\ref{TB:LINAC} with the beam spot and phase spaces shown in Fig.~\ref{FIG:LINAC}. The final two cavities are chirped $-4^{\circ}$ off-crest in order to reduce the \textit{rms} energy spread. At this location, the extremely small transverse normalized \textit{rms} emittance has been achieved.

\begin{table}
\caption{Properties of electron bunch at linac exit.}
\label{TB:LINAC}
\begin{ruledtabular}
\begin{tabular}{lcc}
Parameter & Quantity & Units \\
\hline
kinetic energy & 25. & MeV \\
\textit{rms} energy spread & 3.44 & keV \\
$\epsilon^N_{x,\mathrm{rms}}$ & 0.10 & mm-mrad \\
$\epsilon^N_{y,\mathrm{rms}}$ & 0.13 & mm-mrad \\
$\sigma_x$ & 0.35 & mm \\
$\sigma_y$ & 0.38 & mm \\
$\beta_x$ & 60 & m \\
$\beta_y$ & 54 & m \\
$\alpha_x$ & -2.3 & - \\
$\alpha_y$ & -3.8 & - \\
$\sigma_z$ & 0.67 & mm \\
\end{tabular}
\end{ruledtabular}
\end{table}

\subsection{Emittance Decrease}

It has been noted before that the transverse normalized \textit{rms} emittance of the bunch out of the gun is not necessarily the same out of the linac. In the first iteration of the gun design, there was an increase in emittance after the bunch exited the gun because it was not yet at a sufficient energy to make space charge negligible. In the final design, however, the emittance actually decreases between the gun and linac exits. The final iteration has a greater decrease in emittance and will be examined here to explain the behavior.

\begin{figure}
\includegraphics[width=0.7\columnwidth]{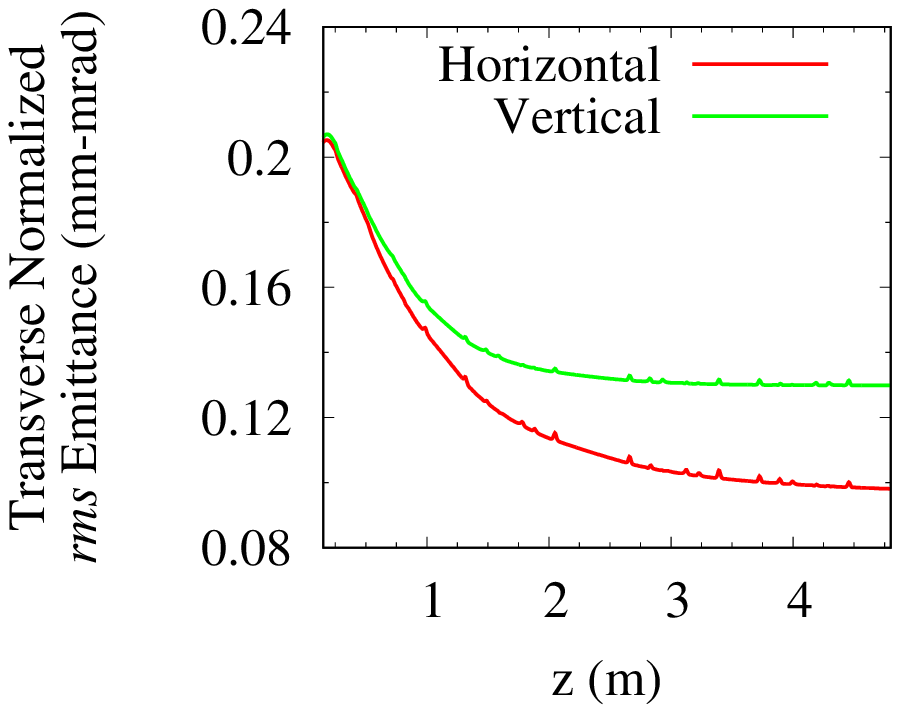}
\includegraphics[width=0.7\columnwidth]{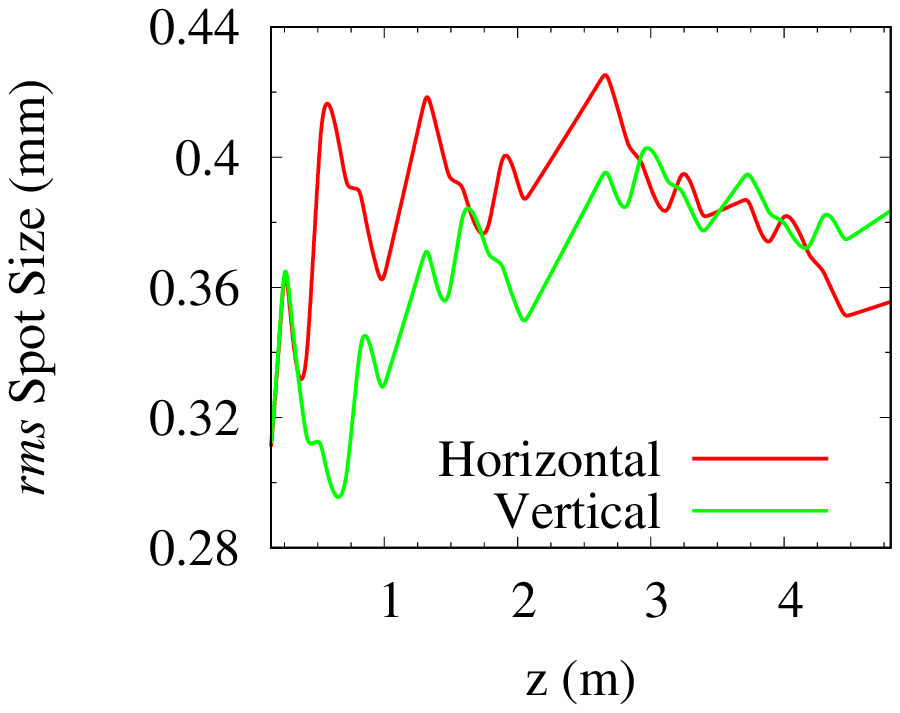}
\caption{Transverse normalized \textit{rms} emittances (top) and spot sizes (bottom) of bunch passing through the linac.}
\label{FIG:EMIT_LINAC_FUNCTIONS}
\end{figure}

The transverse normalized \textit{rms} emittances and \textit{rms} spot sizes of the bunch as it passes through the linac are shown in Fig.~\ref{FIG:EMIT_LINAC_FUNCTIONS}. Both horizontal and vertical emittances decrease through the linac, though the rate of decrease changes with the longitudinal position and which transverse component is being considered. The transverse \textit{rms} sizes of the beam grow rapidly immediately after the bunch exits the gun, but the size increase is limited within the linac.

\begin{figure}
\includegraphics[width=0.7\columnwidth]{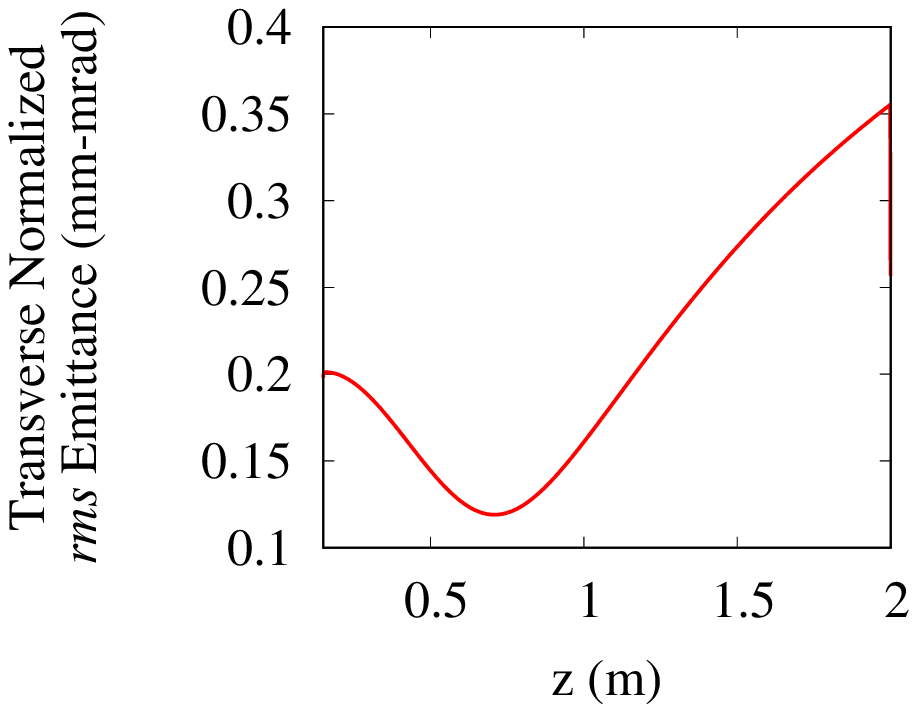}
\includegraphics[width=0.7\columnwidth]{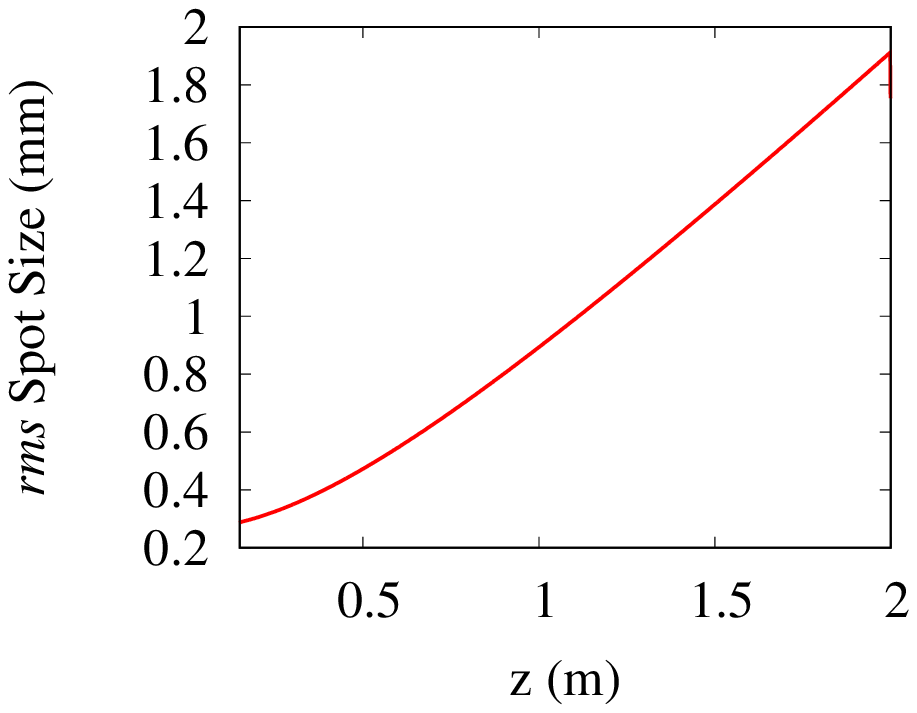}
\caption{Transverse normalized \textit{rms} radial emittance (top) and transverse spot size (bottom) of bunch drifting after gun exit as a function of longitudinal position.}
\label{FIG:EMIT_SIZE_DECREASE}
\end{figure}

\begin{figure*}
\includegraphics[width=0.9\textwidth]{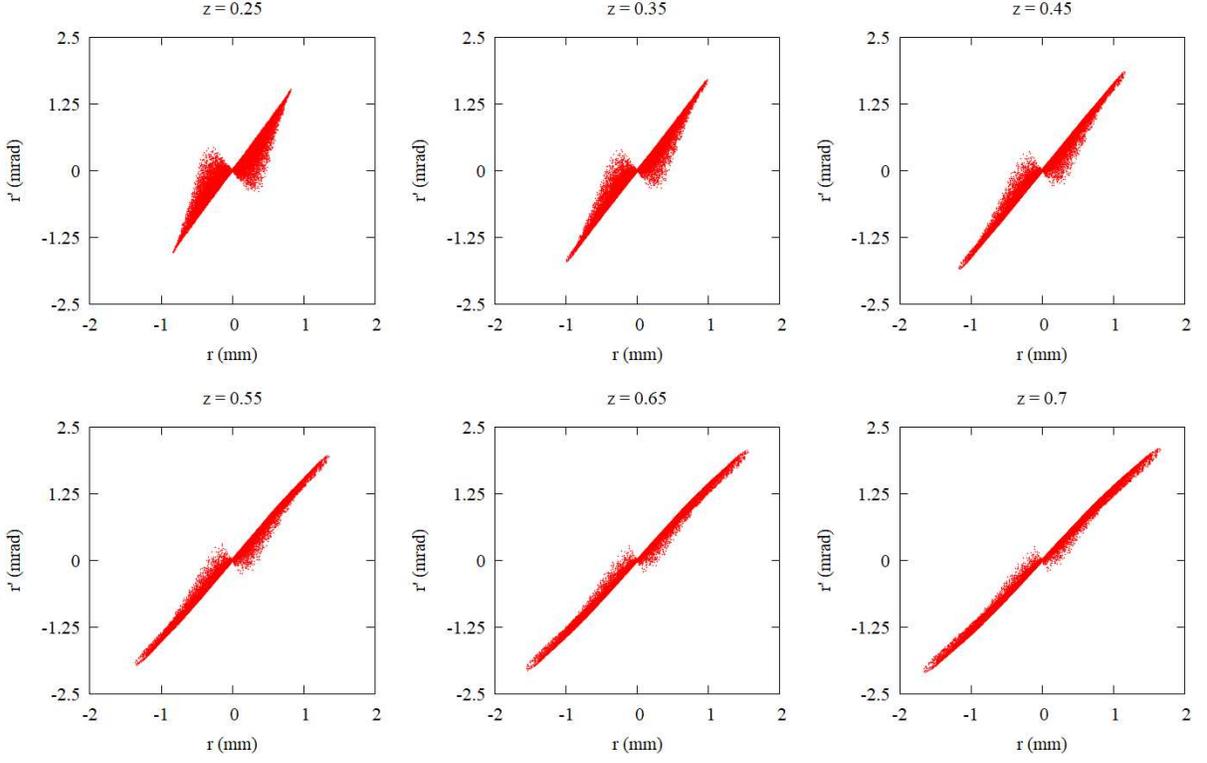}
\caption{Transverse phase spaces of the bunch as it drifts downstream.}
\label{FIG:EMIT_PHASE_SPACES}
\end{figure*}

Using IMPACT-T, it is possible to see the evolution of the bunch after the gun as the beam drifts downstream, without passing through the linac. The transverse normalized \textit{rms} emittance and the spot size of such a drifting bunch are shown as a function of longitudinal position in Fig.~\ref{FIG:EMIT_SIZE_DECREASE}. While the spot size increases as the bunch drifts downstream, the emittance decreases to a minimum at approximately $z = 0.7$~m, before increasing. The transverse phase spaces of the bunch are shown in Fig.~\ref{FIG:EMIT_PHASE_SPACES} at a number of locations after the gun exit, up to and including the minimum at $z = 0.7$~m.

One further aspect of interest is that for the drifting bunch, $\epsilon^N_{\mathrm{rms},r} = 0.12$~mm-mrad at the minimum of $z = 0.7$~m, but at the exit of the linac $\epsilon^N_{\mathrm{rms},x} = 0.10$~mm-mrad and $\epsilon^N_{\mathrm{rms},y} = 0.13$~mm-mrad. So even the average of the two transverse emittances is less than what can be attained if the bunch just drifts after the gun. If the bunch charge of the beam exiting the gun is artificially decreased, the distance to the emittance minimum increases and the emittance minimum decreases. This can be considered analogous to increasing the beam energy without the additional phase space manipulations of passing the beam through the ``quadrupole-like'' spoke cavities.

\begin{figure}
\includegraphics[width=\columnwidth]{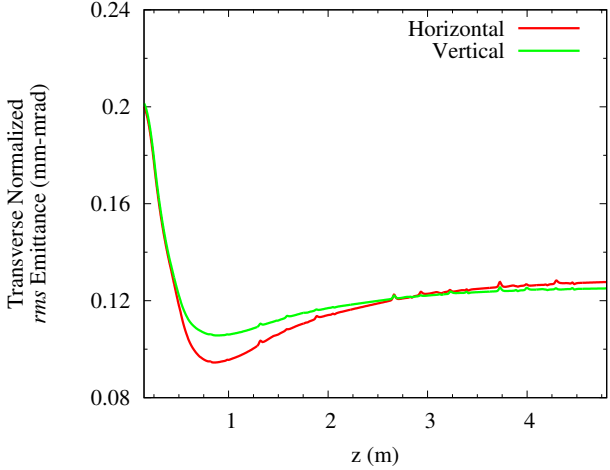}
\caption{Transverse normalized \textit{rms} emittances of the bunch off the cathode tracked through an unoptimized version of the accelerating section as a function of the longitudinal position.}
\label{FIG:EMIT_MIN_LINAC}
\end{figure}

Increasing the energy of the beam does not mean it is impossible for an emittance minimum to occur within the linac; it depends on the bunch exiting the gun. One example of an emittance minimum occurring within the linac is shown in Fig.~\ref{FIG:EMIT_MIN_LINAC}. The figure shows the transverse normalized \textit{rms} emittances of the cathode bunch tracked through an unoptimized version of the accelerating section. While the emittances decrease, after the minimum both increase. At this minimum, $\epsilon^N_{\mathrm{rms},x} = 0.095$~mm-mrad and $\epsilon^N_{\mathrm{rms},y} = 0.11$~mm-mrad, both of which are smaller values, respectively, than those of the bunch exiting the linac. With the increase after the minimum, the bunch exits with $\epsilon^N_{\mathrm{rms},x} = 0.13$~mm-mrad and $\epsilon^N_{\mathrm{rms},y} = 0.13$~mm-mrad, so this is not the best possible system for this initial bunch. Consequently, there is some limit on the rate of emittance decrease for the bunch exiting the gun. If the emittance decreases too rapidly, a minimum occurs within the linac, which leads to the beam quality suffering.

\section{\label{SEC:FOCUSING}Final Focusing}

\begin{figure}
\includegraphics[width=\columnwidth]{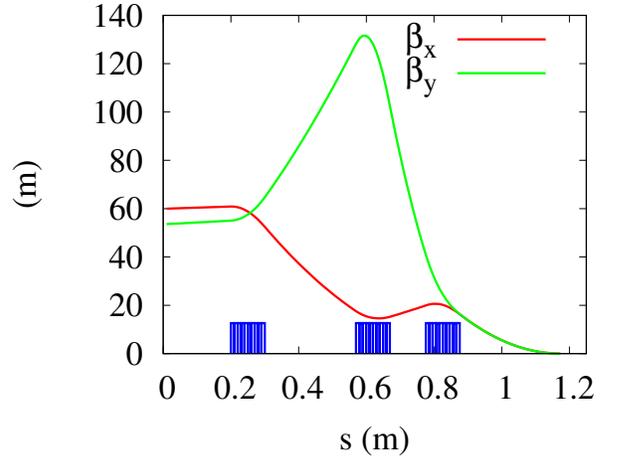}
\caption{$\beta_x$ and $\beta_y$ as a function of $s$ in the final focusing section of the design. The location of the three quadrupoles are positioned along the horizontal axis.}
\label{FIG:FINAL_SHAPE}
\end{figure}

\begin{figure*}
\includegraphics[width=0.7\textwidth]{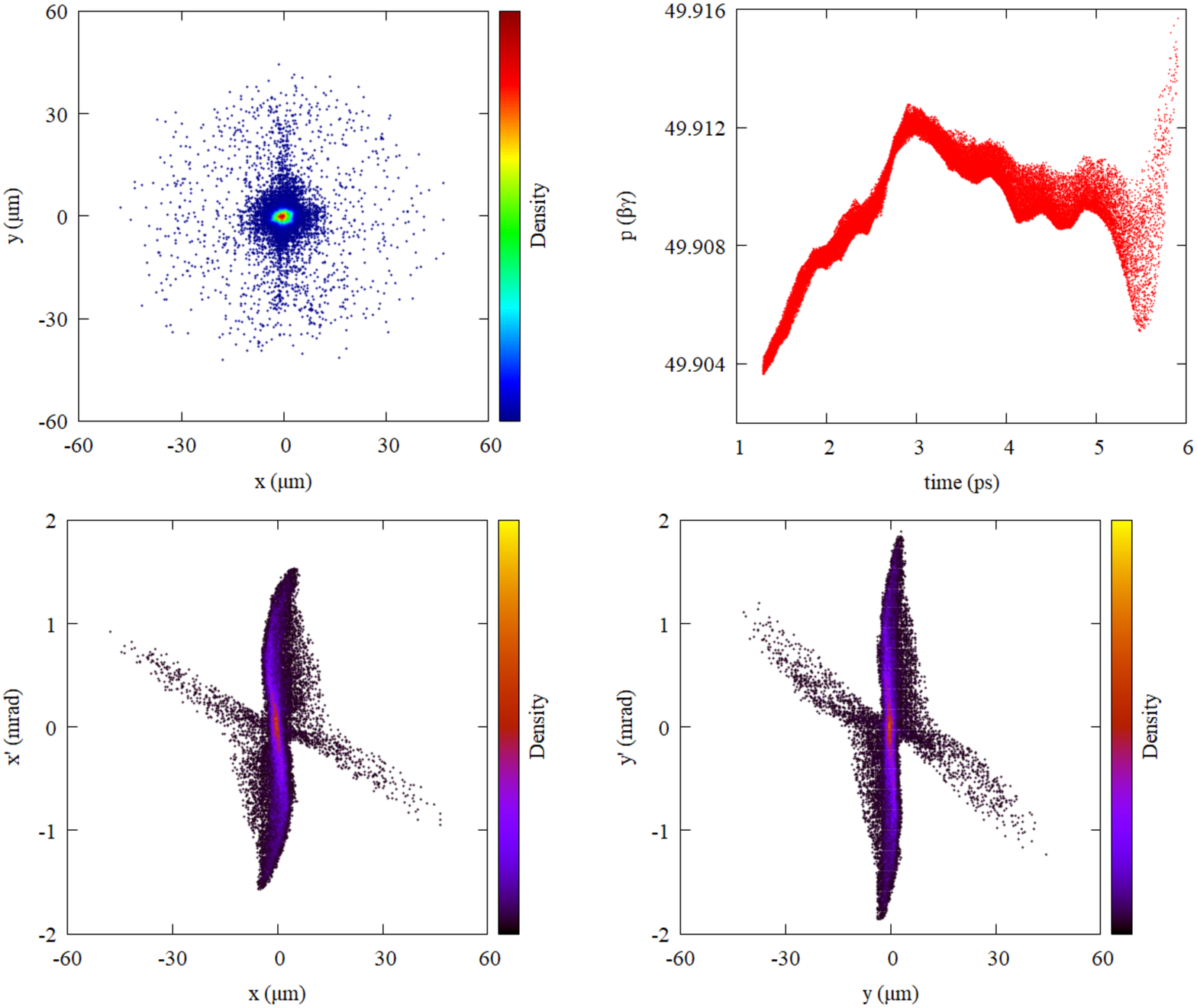}
\caption{Beam spot (top left), longitudinal phase space (top right), horizontal phase space (bottom left), and vertical phase space (bottom right) of the electron bunch at the IP.}
\label{FIG:FINAL_IP}
\end{figure*}

The final focusing section consists of three quadrupoles, with a distance of $\sim$29~cm between the third quadrupole and the IP. Tracking was performed using \texttt{elegant} and the bunch distribution at the exit of the linac~\cite{elegant}. \texttt{Elegant} was used in order to make the optimization easier, but the results were later compared to tracking using a 3D space charge tracking code and found to have negligible differences. Aberrations of quadrupole displacement were included in the sensitivity studies. The value of $\beta_x$ and $\beta_y$ are shown as a function of the beam path $s$ in Fig.~\ref{FIG:FINAL_SHAPE}. Certain aspects of the focusing lattice and the properties of the bunch at the IP are shown in Tables~\ref{TB:FINAL_IP} and~\ref{TB:FINAL_IP_E}, respectively. The beam spot and phase spaces of this beam are shown in Fig.~\ref{FIG:FINAL_IP}. It can be seen that the small emittance is preserved while focusing the electron beam spot size to a few microns.

\begin{table}
\caption{Select magnet properties of the final focusing section.}
\label{TB:FINAL_IP}
\begin{ruledtabular}
\begin{tabular}{lcc}
Parameter & Quantity & Units \\
\hline
Maximum $\beta$ & 132 & m \\
Quadrupole length & 0.1 & m \\
Quadrupole strengths & 1.2 - 3.6 & T/m \\
\textit{rms} energy spread & 3.4 & keV \\
\end{tabular}
\end{ruledtabular}
\end{table}

\begin{table}
\caption{Select properties of the electron beam parameters, both desired and achieved, at the IP.}
\label{TB:FINAL_IP_E}
\begin{ruledtabular}
\begin{tabular}{lccc}
Parameter & Desired & Achieved & Units \\
\hline
$\beta_x$ & 5 & 5.4 & mm \\
$\beta_y$ & 5 & 5.4 & mm \\
$\epsilon^N_{\mathit{rms},x}$ & 0.1 & 0.1 & mm-mrad \\
$\epsilon^N_{\mathit{rms},y}$ & 0.1 & 0.13 & mm-mrad \\
$\sigma_x$ & 3.2 & 3.4 & $\mu$m \\
$\sigma_y$ & 3.2 & 3.8 & $\mu$m \\
$>76\%$ longitudinal& 3 & 3 & ps \\
\hspace{5mm}distribution & & & \\
\textit{rms} energy spread & 7.5 & 3.4 & keV \\
\end{tabular}
\end{ruledtabular}
\end{table}

\begin{figure*}
\includegraphics[width=0.7\textwidth]{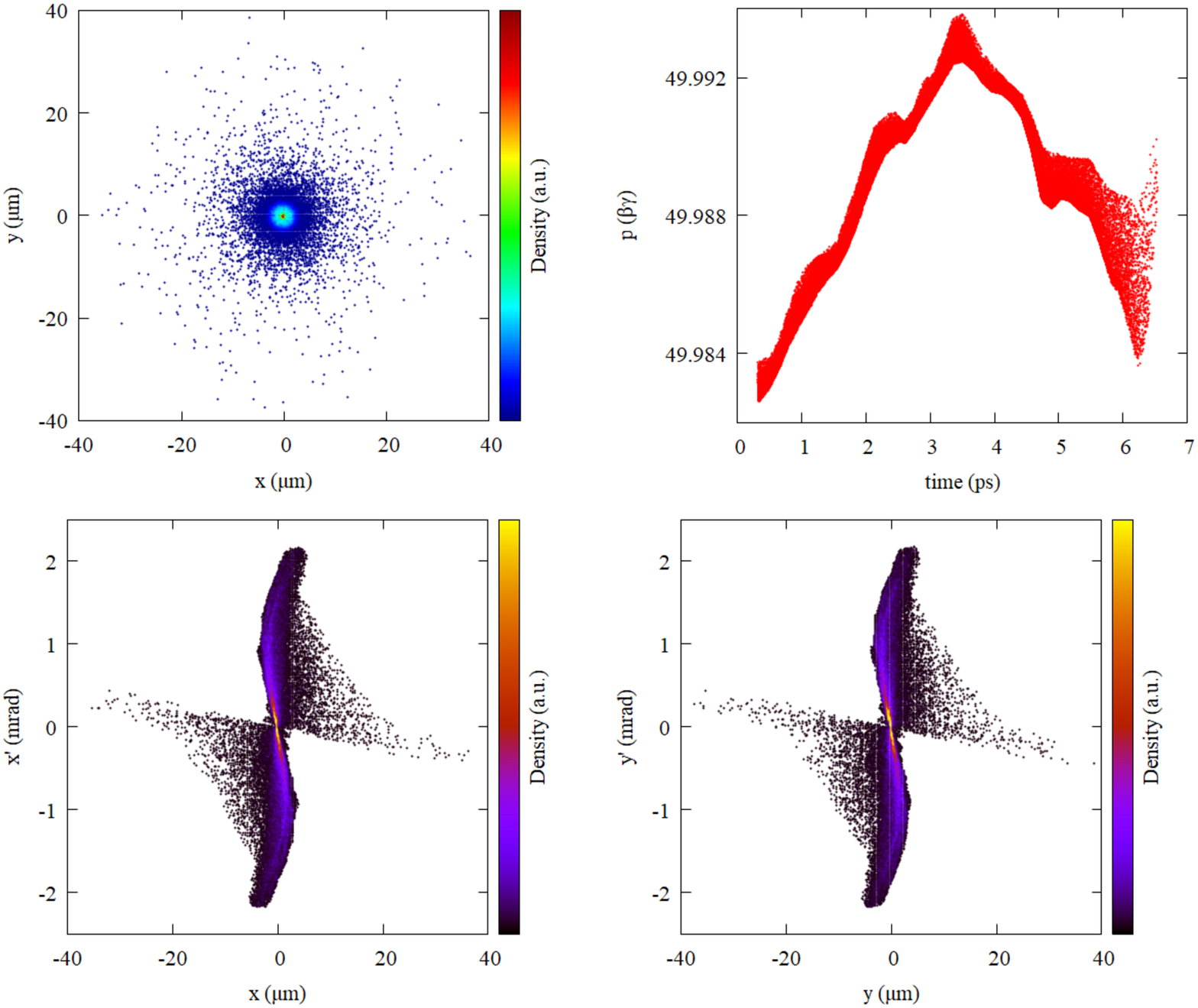}
\caption{Beam spot (top left), longitudinal phase space (top right), horizontal phase space (bottom left), and vertical phase space (bottom right) of the electron bunch at the IP, using elliptical cavities in the linac.}
\label{FIG:ELLIPTICAL_IP}
\end{figure*}

There is an assumption that the use of spoke cavities in the linac instead of traditional transverse magnetic (TM) mode cavities, such as multicell elliptical cavities, has a detrimental effect on the transverse emittance of the beam. In order to examine this idea, each double-spoke cavity in the linac was replaced by a 3-cell elliptical cavity; the EM fields were calculated using Superfish. After using  IMPACT-T to track the bunch through this version of the linac, the beam was focused to a small size. The beam spot and phase spaces of the focused bunch is shown in Fig.~\ref{FIG:ELLIPTICAL_IP}. When compared to the bunch accelerated by the double-spoke cavities, seen in Fig.~\ref{FIG:FINAL_IP}, the beam is highly comparable. The transverse normalized \textit{rms} emittance of the new bunch is $\sim0.11$ mm-mrad, which is the average transverse normalized \textit{rms} emittance of the bunch using spoke cavities. Consequently, there is no beam physics reason that elliptical cavities are the better option for beam acceleration.

\section{\label{SEC:JITTER}Sensitivity Studies}

In order to ascertain the robustness of the design, simulations involving deviations from optimal design parameters were performed. In these simulations, the maximum threshold for each perturbation from the design was determined to be the point when any electron beam parameter value given in Table~\ref{TB:FINAL_IP} changed by 20\%.

The phase and amplitude of each SRF structure was individually varied while holding all other settings constant. The change in phase is given in degrees, while the change in amplitude is given in percentage of the design value. The thresholds are reported in Table~\ref{TB:JITTER}. The limiting electron beam parameter is the vertical \textit{rms} size when the amplitude of the cavities is varied. In all other cases, the limiting parameter is the \textit{rms} energy spread.

\begin{table}
\caption{The amplitude and phase perturbation from design for each SRF structure at which some electron beam parameter changes $\sim$20\% at the IP.}
\label{TB:JITTER}
\begin{ruledtabular}
\begin{tabular}{lc}
Varied Parameter and Structure & Threshold\\
\hline
Amplitude of Gun & -2.0\%\\
 & +0.6\% \\
Amplitude of All Cavities & -1.0\% \\
 & +0.8\% \\
Phase of Gun & -7.2$^{\circ}$ \\
 & +1.2$^{\circ}$ \\
Phase of All Cavities & -1.2$^{\circ}$ \\
 & +1.2$^{\circ}$ \\
\end{tabular}
\end{ruledtabular}
\end{table}

Systematic perturbations were also evaluated for the coordinates of both the linac cavities and the magnets in the final focusing section, separately. In either case, each element was randomly attributed some amount of misalignment in each of the three Cartesian directions. The maximum possible misalignment is the threshold. For the translational (transverse and longitudinal) misalignment in the linac cavities, the threshold is 500~$\mu$m, with the limiting electron beam parameter being the \textit{rms} energy spread. For the translational misalignment of the three quadrupole magnets, the threshold is 300~$\mu$m, with the limiting parameter being the vertical size~\cite{Diss}.

\section{\label{SEC:LASER}Incident Laser}

Inverse Compton Light Sources require both an electron beam and an incident laser, the latter of which has been neglected thus far. The design of the appropriate laser is beyond the scope of this article, but the desired properties are provided in Table~\ref{TB:LASER_IP_DESIRED}. While a laser with a circulating power of 1~MW is called for, such a laser does not currently exist. Current consensus of those within that field is that such a laser is feasible, but until now there has not been a need for it to be developed. At present, high average power lasers currently constructed have a power of $\sim$100~kW. Using a laser with this circulating power would decrease the flux and brightness of the anticipated X-ray beam by an order of magnitude.

\section{\label{SEC:XRAY}X-ray Source}

By using the parameter values in Tables~\ref{TB:FINAL_IP} and \ref{TB:LASER_IP_DESIRED} in the formulae presented in Sec.\ref{SEC:XRAY} and \ref{SEC:COMPTON_SCATTERING}, it is possible to estimate the X-ray beam parameters of the light source. These parameters are presented in Table~\ref{TB:XRAY_FORMULAE}, assuming Gaussian laser and beam spots. However, the electron distribution at the interaction point is not Gaussian, bringing the validity of the results into question.

\begin{table*}
\caption{Estimated X-ray performance assuming design electron beam attained at IP, compared to desired parameters.}
\label{TB:XRAY_FORMULAE}
\begin{ruledtabular}
\begin{tabular}{lccccc}
Parameter & \multicolumn{4}{c}{Laser Spot Size ($\mu$m)} & Units \\
\cline{2-5}
 & \multicolumn{2}{c}{3.2} & \multicolumn{2}{c}{12} & \\
 & Desired & Achieved & Desired & Achieved & \\
\hline
X-ray energy & 12 & 12 & 12 & 12 & keV \\
$N_{\gamma}$ & $1.6 \times 10^6$ & $1.4 \times 10^6$ & $2.1 \times 10^5$ & $2.1 \times 10^5$ & photons/bunch \\
Flux & $1.6 \times 10^{14}$ & $1.4 \times 10^{14}$ & $2.1 \times 10^{13}$ & $2.1 \times 10^{13}$ & ph/s \\
Flux in 0.1\% BW & $2.4 \times 10^{11}$ & $2.1 \times 10^{11}$ & $3.2 \times 10^{10}$ & $3.1 \times 10^{10}$ & ph/(s-0.1\%BW) \\
Average Brilliance & $3.0 \times 10^{15}$ & $2.2 \times 10^{15}$ & $2.1 \times 10^{14}$ & $1.6 \times 10^{14}$ & ph/(s-mm\textsuperscript{2}\\
& & & & & -mrad\textsuperscript{2}-0.1\%BW) \\
\end{tabular}
\end{ruledtabular}
\end{table*}

\begin{figure*}
\includegraphics[width=0.95\columnwidth]{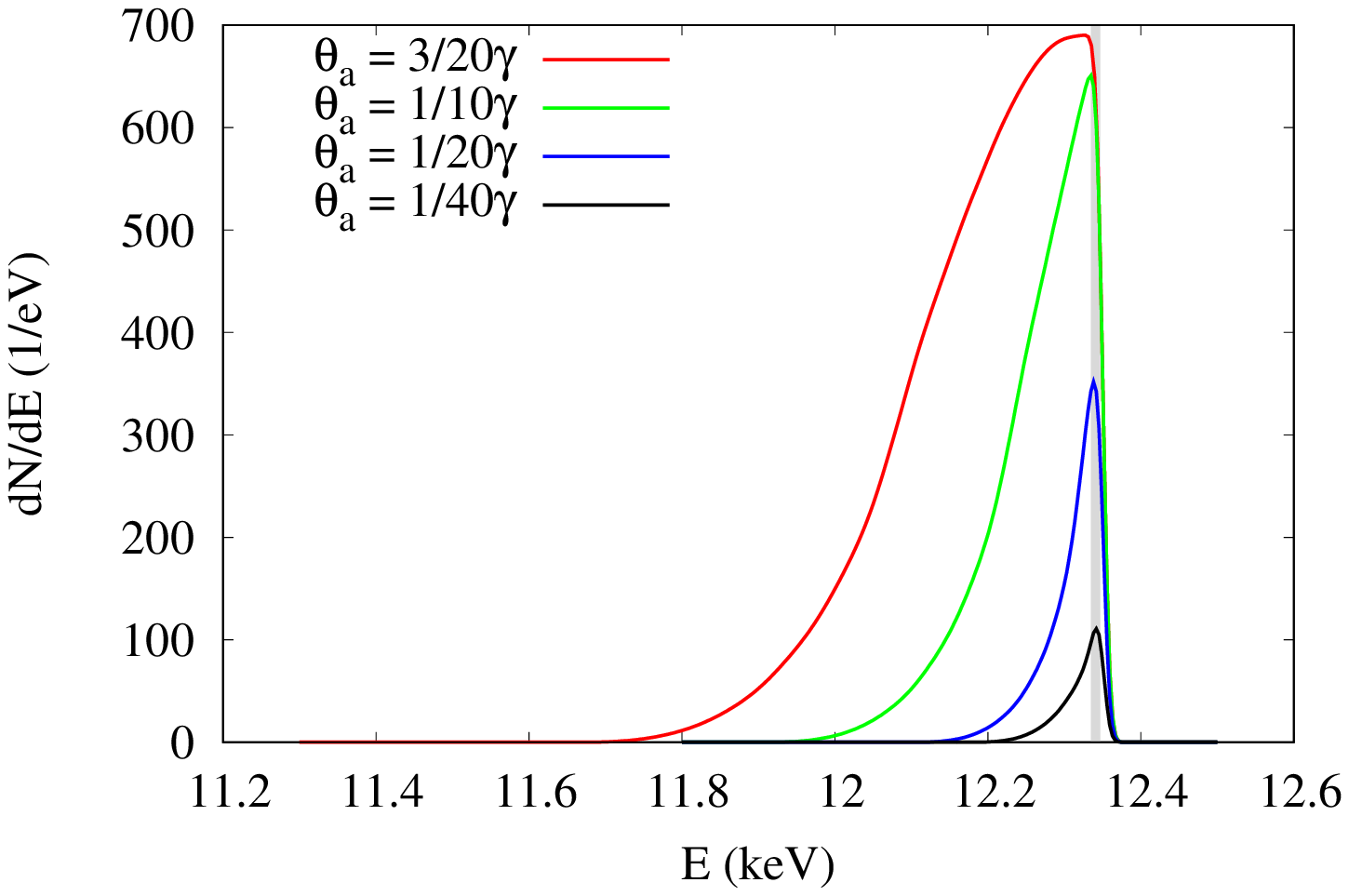}
\includegraphics[width=0.95\columnwidth]{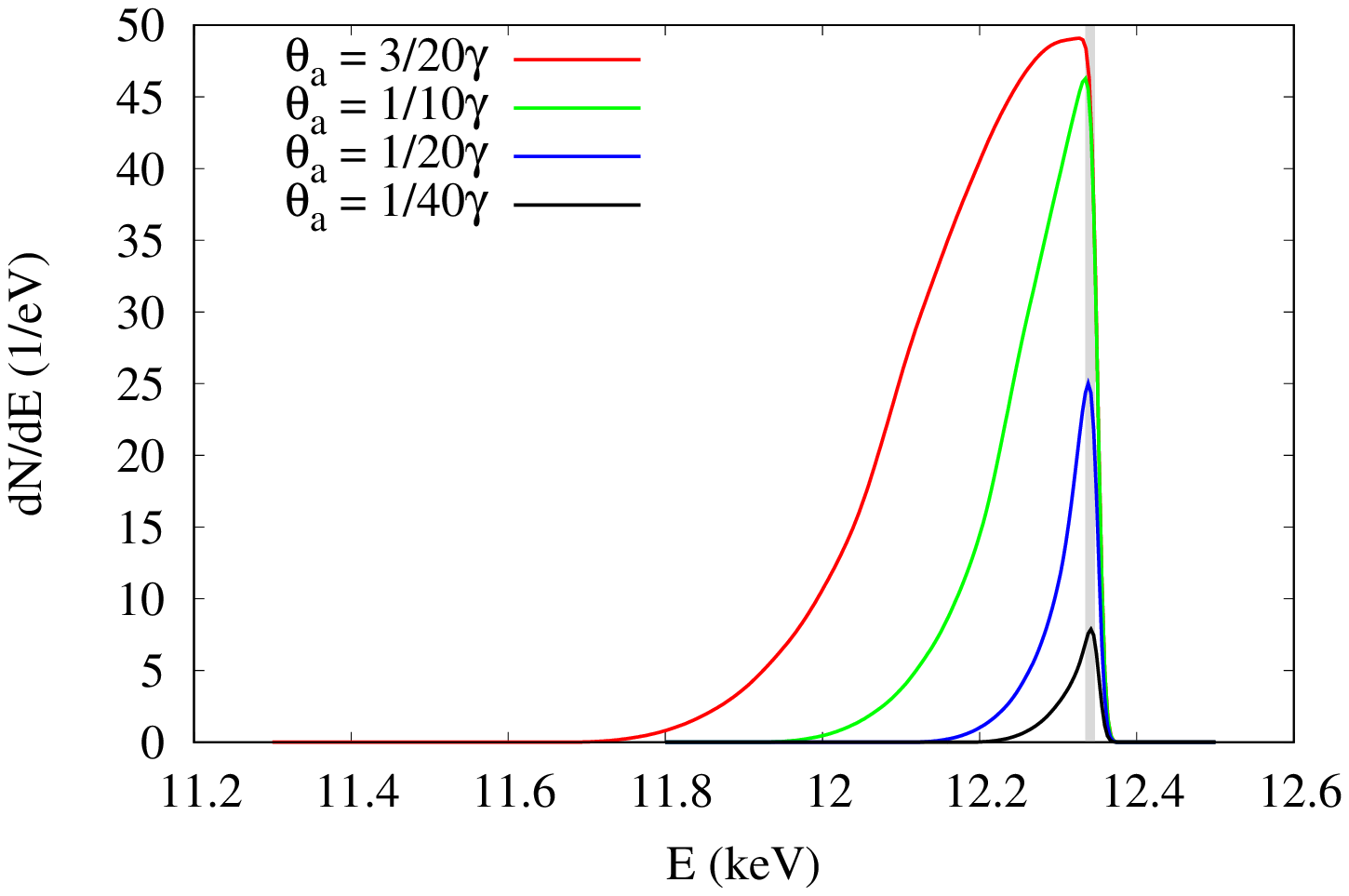}
\caption{Number spectra for different apertures generated using 4,000 particles for a 3.2~$\mu$m laser spot size (left) and a 12~$\mu$m laser spot size (right). Grey box indicates 0.1\% bandwidth. Suggested apertures for brilliance calculation only.}
\label{FIG:XRAY_SPECTRUM}
\end{figure*}

Fortunately, Compton scattering calculations have been made recently which utilize the electron beam distribution, not just beam parameters. Using these methods, the calculations of the X-ray light source parameters verify that the non-Gaussian distribution does not significantly impact the anticipated brilliance~\cite{PRAB_16}. In the previous paper \cite{PRAB_16}, we used 
\begin{equation}
\mathcal{B}_{\mathit{p}} = \lim_{\theta_a\rightarrow0}\frac{\mathcal{S}_{0.1\%}}{2\pi^2\sigma_e^2\theta_a^2}
\end{equation}
to calculate the pin-hole brilliance of the X-ray beam $\mathcal{B}_{\mathit{p}}$, where $\mathcal{S}_{0.1\%}$ is the number of photons per second in a 0.1\% bandwidth transmitted through the aperture, which is calculated from the code. However, applying the same reasoning given in Sec.\ref{SEC:XRAY_PARAS}, this formula becomes
\begin{equation}
\mathcal{B}_{\mathit{p}} = \lim_{\theta_a\rightarrow0}\frac{\mathcal{S}_{0.1\%}}{2\pi^2\sigma_{\gamma}^2\theta_a^2}.
\end{equation}
Table~\ref{TB:XRAY_ERIK} contains the estimated X-ray beam parameters when calculated in this manner, while the full spectrum of the produced X-rays is shown in Fig.~\ref{FIG:XRAY_SPECTRUM}. 

\begin{table}
\caption{X-ray performance of the designs attained by numerical simulation with an aperture of 1/40$\gamma$. Suggested aperture for brilliance calculation only.}
\label{TB:XRAY_ERIK}
\begin{ruledtabular}
\begin{tabular}{lccc}
Parameter & \multicolumn{2}{c}{Laser Spot ($\mu$m)} & Units \\
 & 3.2 & 12 & \\
\hline
X-ray energy & 12.3 & 12.3 &  keV \\
$N_{0.1\%}$ & 1230 & 92.4 & ph/0.1\%BW\\
$\mathcal{S}_{0.1\%}$ & $1.23 \times 10^{11}$ & $9.24 \times 10^{9}$ & ph/(s-0.1\%BW) \\
$\mathcal{B}_p$ & $4.61 \times 10^{15}$ & $1.58 \times 10^{14}$ & ph/(s-mm\textsuperscript{2} \\
 & & & -mrad\textsuperscript{2}-0.1\%BW)\\
\end{tabular}
\end{ruledtabular}
\end{table}

The energy and average brilliance in Tables~\ref{TB:XRAY_FORMULAE} and \ref{TB:XRAY_ERIK} show excellent agreement for the 12~$\mu$m case. The flux into a 0.1\% bandwidth does not, which is expected because this parameter is dependent on the aperture angle of the interaction enclosure. Calculation results clearly demonstrate that the flux increases with the aperture angle, and the results are being reported for a small opening. The factor of two increase between the calculated brilliance and the pin-hole brilliance for the 3.2~$\mu$m case is expected - the calculation code makes the assumption that every photon of the scattering laser sees the same scattering potential, which is most valid when the laser spot size is much greater than the electron beam spot size. As that assumption is not valid for the 3.2~$\mu$m case, the code overestimates the anticipated pin-hole brilliance. 

\section{\label{SEC:FURTHER}Further Work}

While we have presented here a preliminary design for a high-brilliance, high-flux inverse Compton light source much work remains to be done (analytical, numerical simulation, and experimental) before such a source could be built.  In particular, it may be that the choices of frequency and geometry are not optimal and may be revised after further study or advances is the SRF technology.  In this section we outline the major items that would require further R\&D activities.

\subsection{SRF Gun}

A number of simulation studies need to be conducted before the SRF gun is built and commissioned - primarily multipacting and mechanical. If geometry alterations are deemed necessary, another optimization based on simulation beam dynamics results may be necessary, including these studies.

Once these studies are completed and the geometry is finalized, the gun will need to be built and commissioned. At this point, experiments can be performed to demonstrate that the transverse normalized \textit{rms} emittance behavior as the beam drifts after the gun is as expected, supporting the idea that appropriate gun geometry can provide all necessary emittance compensation.

Integration of a photo-cathode and an SRF gun presents a technical challenge which is under investigation in a number of institutions.

\subsection{Beam Physics}

Further sensitivity studies are called for, especially if the gun geometry needs to be altered to avoid significant obstacles in multipacting, mechanical stability, or thermal breakdown. Additionally, simulations examining potential wakefields are desired.  All these simulations will need to include potential deviations from an ideal design (physical misalignment, errors of amplitude and phase, etc). 

\subsection{SRF Linac}

The design presented here is based on a particular choice of frequency and geometry in order to achieve a balance between size (capital cost) and operating cost.  This choice may not be optimal and, in particular, the operating cost would still be higher than desired based on the surface resistance assumed in Table~\ref{TB:DOUBLESPOKE}.  However recent progress in SRF R\&D shows potential for a substantial reduction of power dissipation in SRF cavities.  Nitrogen doping \cite{N2doping} and infusion \cite{N2infusion} during heat treatment have shown large reduction of those losses at 2 K and higher frequency.  Achieving similar results at 4.2 K and 500 MHz would validate our choices.  On a longer term, Nb\textsubscript{3}Sn \cite{Nb3Sn} could offer dramatic reduction in cryogenic losses at 4.2 K and would even allow operation at higher frequency; such an advance would lead us to revisit our choices as multi-cell TM-type cavities operating around 650 MHz would be attractive.  They may even be able to operate without a refrigerator, using instead cryocoolers \cite{compact} if vibrations that such systems often generated can be managed.

Eventually, several prototype cavities will need to be built and tested, and all the processes needed to achieve performance (chemistry, heat treatment, cleaning, etc) will have to be developed and demonstrated.

Performance of this light source is contingent upon achieving and maintaining a very small emittance. This will put challenging constraints on the design of the cryomodule and the Low Level RF Control system.

\subsection{Incident Laser}

As previously mentioned, a laser with the desired properties of either spot size does not currently exist. While current technology might suffice in providing an X-ray beam at least as good as any other compact ICLS, this design has the capacity to surpass that threshold. Consequently, such a laser must be designed and commissioned, before the project presented here is commissioned.

The benefits of developing a better incident laser do not stop with this project, however. Other compact ICLS projects, proposed or existing, can see an improvement in the quality of the X-ray beam they deliver. This approach may be one method which will see such development funded. Additionally, other applications for a more powerful laser do exist.

\subsection{Overall}

A complete design should be produced, including all necessary components - klystrons, cathode drive laser, refrigeration support, beam dump, etc. The commissioning process itself is likely to be divided into two main stages - the electron beam and the X-ray beam. Initially, the electron beam will need to be produced at the intended interaction point, with the intended properties. Afterwards, the incident scattering laser will be installed, with the appropriate beam transport to allow the produced X-ray beam to travel to the users and the electron beam to travel to the beam dump.

\section{\label{SEC:CONCLUSION}Summary}

The Compact ICLS design presented here would improve on all other compact sources to date, producing an X-ray beam of quality which is closer than ever to being comparable to beams produced at large-scale facilities. This is made possible by using cw superconducting RF to accelerate the beam before it is focused to the interaction point. At the interaction point, the electron beam has a small spot size and small transverse normalized \textit{rms} emittance, which correspondingly result in an X-ray beam with high flux and brilliance. The ultra-low emittance is made feasible by a low bunch charge, with a high repetition rate so the X-ray flux is not adversely affected.

This design achieves an electron bunch which generates an X-ray beam unmatched in quality by other Compact ICLS designs. These desired electron beam parameters are achieved by utilizing a number of different techniques. The most effective technique was the emittance compensation by RF focusing. By altering the geometry of the gun to provide the correct RF focusing for a given bunch, it is possible to produce bunches with low normalized transverse \textit{rms} emittances. Taken together with the low bunch charge, the achieved transverse emittances are sufficiently small. Choosing the correct bunch length off the cathode is necessary, in order to produce a bunch exiting the linac which does not need compression, but is still long enough that the transverse space charge effects can be compensated for by the RF focusing provided by the gun geometry. Another beneficial technique is taking advantage of the quadrupole-like behavior of the double-spoke cavities which comprise the linac in order to produce a fairly round beam at the exit of the linac. An approximately round beam at the exit to the linac allows for the bunch to be easily focused down to a small spot size on the order of a few microns.

While the incident laser remains an incomplete component, its design should not be an obstacle. Further work outside of this aspect includes further optimization to improve on the current X-ray parameters or altering the design for different functions - by increasing the X-ray energy, for example.

\begin{acknowledgments}
The authors would like to thank Rocio Olave, Karim Hern\'{a}ndez-Chah\'{i}n, Subashini De Silva, Christopher Hopper, Randika Gamage, and Todd Satogata for earlier contributions to this project, and HyeKyoung Park for a discussion of cavity surface treatments. This material is based on work supported by the U. S. Department of Energy, Office of Science, Office of Nuclear Physics and Office of Basic Energy Science; and by the National Science Foundation.  K.E.D and J.R.D. were supported at ODU by DOE Office of Nuclear Physics award  No. DE-SC00004094.  B.T. was supported at ODU by NSF Award 1535641.  G.A.K was supported at Jefferson Lab by contract DE-AC05-06OR23177; additional support was provided by DOE Office of Nuclear Physics Award No. DE-SC004094 and Basic Energy Sciences Award No. JLAB-BES11-01.  This research used resources of the National Energy Research Scientific Center, which is supported by the Office of Science of the U.S. DOE under Contract No. DE-AC02-05CH11231.
\end{acknowledgments}

\bibliography{PhysRevDiss}

\end{document}